\documentclass[longauth]{aa}

\usepackage{graphicx}
\usepackage{natbib}
\usepackage{scalerel}

\usepackage[table]{xcolor}

\usepackage{txfonts}

\usepackage[pdfencoding=auto,psdextra]{hyperref}
\hypersetup{
    colorlinks=true,
    linkcolor=blue,
    filecolor=magenta,
    urlcolor=blue,
    citecolor=blue
}
\defcitealias{EROData}{C25}
\defcitealias{Q1-SP007}{K26}
\defcitealias{Q1-SP008}{GM26}

\makeatletter
\renewcommand*\aa@pageof{, page \thepage{} of \pageref*{LastPage}}
\makeatother

\usepackage[utf8]{inputenc}

\usepackage[switch, modulo]{lineno}

\renewcommand{\linenumbers}{}

\usepackage{euclid}

\usepackage{siunitx}

\providecommand{\qty}[2]{\SI{#1}{#2}}
\providecommand{\unit}[1]{\si{#1}}

\newcommand{\Msun}{M_{\sun}}

\newcommand{\swarp}{\texttt{SWarp}}
\newcommand{\YJH}{$YJH_\sfont{E}$\xspace}

\newcommand\reffig[1]{Fig.~\ref{fig:#1}}
\newcommand\refFig[1]{Figure~\ref{fig:#1}}
\newcommand\refsec[1]{Sect.~\ref{sc:#1}}
\newcommand\refSec[1]{Section~\ref{sc:#1}}
\newcommand\refApp[1]{Appendix~\ref{ap:#1}}
\newcommand\refapp[1]{Sect.~\ref{ap:#1}}

\newcommand{\snr}{\ensuremath{\mathrm{S}/\mathrm{N}}\xspace}

\newcommand{\changed}[1]{#1}

\begin{document}

\title{Euclid Quick Data Release (Q1)}
\subtitle{Data processing for intracluster light measurements}

\newcommand{\orcid}[1]{}
\author{Euclid Collaboration: S.~P.~Bamford\orcid{0000-0001-7821-7195}\inst{\ref{aff1}}
\and H.~Gao\orcid{0000-0003-1015-5367}\thanks{\email{icl-group@nottingham.ac.uk}}\inst{\ref{aff1}}
\and N.~A.~Hatch\orcid{0000-0001-5600-0534}\inst{\ref{aff1}}
\and T.~Kolcu\orcid{0000-0002-3365-9210}\inst{\ref{aff1}}
\and A.~Fernandez\orcid{0009-0001-4673-526X}\inst{\ref{aff1}}
\and G.~Martin\orcid{0000-0003-2939-8668}\inst{\ref{aff1}}
\and Y.~M.~Bah\'{e}\orcid{0000-0002-3196-5126}\inst{\ref{aff1},\ref{aff2}}
\and J.~B.~Golden-Marx\orcid{0000-0002-6394-045X}\inst{\ref{aff1}}
\and A.~H.~Gonzalez\orcid{0000-0002-0933-8601}\inst{\ref{aff3}}
\and H.~Gully\orcid{0000-0001-7616-7278}\inst{\ref{aff1}}
\and M.~Montes\orcid{0000-0001-7847-0393}\inst{\ref{aff4}}
\and M.~Urbano\orcid{0000-0001-5640-0650}\inst{\ref{aff1},\ref{aff5}}
\and M.~Kluge\orcid{0000-0002-9618-2552}\inst{\ref{aff6}}
\and P.-A.~Duc\orcid{0000-0003-3343-6284}\inst{\ref{aff5}}
\and D.~Scott\orcid{0000-0002-6878-9840}\inst{\ref{aff7}}
\and B.~Altieri\orcid{0000-0003-3936-0284}\inst{\ref{aff8}}
\and S.~Andreon\orcid{0000-0002-2041-8784}\inst{\ref{aff9}}
\and N.~Auricchio\orcid{0000-0003-4444-8651}\inst{\ref{aff10}}
\and C.~Baccigalupi\orcid{0000-0002-8211-1630}\inst{\ref{aff11},\ref{aff12},\ref{aff13},\ref{aff14}}
\and M.~Baldi\orcid{0000-0003-4145-1943}\inst{\ref{aff15},\ref{aff10},\ref{aff16}}
\and S.~Bardelli\orcid{0000-0002-8900-0298}\inst{\ref{aff10}}
\and P.~Battaglia\orcid{0000-0002-7337-5909}\inst{\ref{aff10}}
\and A.~Biviano\orcid{0000-0002-0857-0732}\inst{\ref{aff12},\ref{aff11}}
\and M.~Bolzonella\orcid{0000-0003-3278-4607}\inst{\ref{aff10}}
\and E.~Branchini\orcid{0000-0002-0808-6908}\inst{\ref{aff17},\ref{aff18},\ref{aff9}}
\and M.~Brescia\orcid{0000-0001-9506-5680}\inst{\ref{aff19},\ref{aff20}}
\and S.~Camera\orcid{0000-0003-3399-3574}\inst{\ref{aff21},\ref{aff22},\ref{aff23}}
\and V.~Capobianco\orcid{0000-0002-3309-7692}\inst{\ref{aff23}}
\and C.~Carbone\orcid{0000-0003-0125-3563}\inst{\ref{aff24}}
\and V.~F.~Cardone\inst{\ref{aff25},\ref{aff26}}
\and J.~Carretero\orcid{0000-0002-3130-0204}\inst{\ref{aff27},\ref{aff28}}
\and M.~Castellano\orcid{0000-0001-9875-8263}\inst{\ref{aff25}}
\and G.~Castignani\orcid{0000-0001-6831-0687}\inst{\ref{aff10}}
\and S.~Cavuoti\orcid{0000-0002-3787-4196}\inst{\ref{aff20},\ref{aff29}}
\and A.~Cimatti\inst{\ref{aff30}}
\and C.~Colodro-Conde\inst{\ref{aff31}}
\and G.~Congedo\orcid{0000-0003-2508-0046}\inst{\ref{aff32}}
\and C.~J.~Conselice\orcid{0000-0003-1949-7638}\inst{\ref{aff33}}
\and L.~Conversi\orcid{0000-0002-6710-8476}\inst{\ref{aff34},\ref{aff8}}
\and Y.~Copin\orcid{0000-0002-5317-7518}\inst{\ref{aff35}}
\and F.~Courbin\orcid{0000-0003-0758-6510}\inst{\ref{aff36},\ref{aff37},\ref{aff38}}
\and H.~M.~Courtois\orcid{0000-0003-0509-1776}\inst{\ref{aff39}}
\and M.~Cropper\orcid{0000-0003-4571-9468}\inst{\ref{aff40}}
\and J.-C.~Cuillandre\orcid{0000-0002-3263-8645}\inst{\ref{aff41}}
\and H.~Degaudenzi\orcid{0000-0002-5887-6799}\inst{\ref{aff42}}
\and G.~De~Lucia\orcid{0000-0002-6220-9104}\inst{\ref{aff12}}
\and H.~Dole\orcid{0000-0002-9767-3839}\inst{\ref{aff43}}
\and F.~Dubath\orcid{0000-0002-6533-2810}\inst{\ref{aff42}}
\and X.~Dupac\inst{\ref{aff8}}
\and S.~Escoffier\orcid{0000-0002-2847-7498}\inst{\ref{aff44}}
\and M.~Farina\orcid{0000-0002-3089-7846}\inst{\ref{aff45}}
\and R.~Farinelli\inst{\ref{aff10}}
\and F.~Faustini\orcid{0000-0001-6274-5145}\inst{\ref{aff25},\ref{aff46}}
\and S.~Ferriol\inst{\ref{aff35}}
\and P.~Fosalba\orcid{0000-0002-1510-5214}\inst{\ref{aff47},\ref{aff4}}
\and S.~Fotopoulou\orcid{0000-0002-9686-254X}\inst{\ref{aff48}}
\and M.~Frailis\orcid{0000-0002-7400-2135}\inst{\ref{aff12}}
\and E.~Franceschi\orcid{0000-0002-0585-6591}\inst{\ref{aff10}}
\and M.~Fumana\orcid{0000-0001-6787-5950}\inst{\ref{aff24}}
\and L.~Gabarra\orcid{0000-0002-8486-8856}\inst{\ref{aff49}}
\and S.~Galeotta\orcid{0000-0002-3748-5115}\inst{\ref{aff12}}
\and K.~George\orcid{0000-0002-1734-8455}\inst{\ref{aff50}}
\and B.~Gillis\orcid{0000-0002-4478-1270}\inst{\ref{aff32}}
\and C.~Giocoli\orcid{0000-0002-9590-7961}\inst{\ref{aff10},\ref{aff16}}
\and J.~Gracia-Carpio\orcid{0000-0003-4689-3134}\inst{\ref{aff6}}
\and A.~Grazian\orcid{0000-0002-5688-0663}\inst{\ref{aff51}}
\and F.~Grupp\inst{\ref{aff6},\ref{aff52}}
\and W.~G.~Hartley\orcid{0000-0002-4563-5101}\inst{\ref{aff42}}
\and S.~V.~H.~Haugan\orcid{0000-0001-9648-7260}\inst{\ref{aff53}}
\and H.~Hoekstra\orcid{0000-0002-0641-3231}\inst{\ref{aff54}}
\and W.~Holmes\orcid{0009-0007-8554-4646}\inst{\ref{aff55}}
\and I.~M.~Hook\orcid{0000-0002-2960-978X}\inst{\ref{aff56}}
\and F.~Hormuth\inst{\ref{aff57}}
\and A.~Hornstrup\orcid{0000-0002-3363-0936}\inst{\ref{aff58},\ref{aff59}}
\and K.~Jahnke\orcid{0000-0003-3804-2137}\inst{\ref{aff60}}
\and M.~Jhabvala\inst{\ref{aff61}}
\and B.~Joachimi\orcid{0000-0001-7494-1303}\inst{\ref{aff62}}
\and S.~Kermiche\orcid{0000-0002-0302-5735}\inst{\ref{aff44}}
\and A.~Kiessling\orcid{0000-0002-2590-1273}\inst{\ref{aff55}}
\and B.~Kubik\orcid{0009-0006-5823-4880}\inst{\ref{aff35}}
\and M.~K\"ummel\orcid{0000-0003-2791-2117}\inst{\ref{aff52}}
\and M.~Kunz\orcid{0000-0002-3052-7394}\inst{\ref{aff63}}
\and H.~Kurki-Suonio\orcid{0000-0002-4618-3063}\inst{\ref{aff64},\ref{aff65}}
\and A.~M.~C.~Le~Brun\orcid{0000-0002-0936-4594}\inst{\ref{aff66}}
\and S.~Ligori\orcid{0000-0003-4172-4606}\inst{\ref{aff23}}
\and P.~B.~Lilje\orcid{0000-0003-4324-7794}\inst{\ref{aff53}}
\and V.~Lindholm\orcid{0000-0003-2317-5471}\inst{\ref{aff64},\ref{aff65}}
\and I.~Lloro\orcid{0000-0001-5966-1434}\inst{\ref{aff67}}
\and M.~Magliocchetti\orcid{0000-0001-9158-4838}\inst{\ref{aff45}}
\and G.~Mainetti\orcid{0000-0003-2384-2377}\inst{\ref{aff68}}
\and O.~Mansutti\orcid{0000-0001-5758-4658}\inst{\ref{aff12}}
\and O.~Marggraf\orcid{0000-0001-7242-3852}\inst{\ref{aff69}}
\and M.~Martinelli\orcid{0000-0002-6943-7732}\inst{\ref{aff25},\ref{aff26}}
\and N.~Martinet\orcid{0000-0003-2786-7790}\inst{\ref{aff70}}
\and F.~Marulli\orcid{0000-0002-8850-0303}\inst{\ref{aff71},\ref{aff10},\ref{aff16}}
\and R.~J.~Massey\orcid{0000-0002-6085-3780}\inst{\ref{aff72}}
\and N.~Mauri\orcid{0000-0001-8196-1548}\inst{\ref{aff30},\ref{aff16}}
\and E.~Medinaceli\orcid{0000-0002-4040-7783}\inst{\ref{aff10}}
\and S.~Mei\orcid{0000-0002-2849-559X}\inst{\ref{aff73},\ref{aff74}}
\and M.~Melchior\inst{\ref{aff75}}
\and M.~Meneghetti\orcid{0000-0003-1225-7084}\inst{\ref{aff10},\ref{aff16}}
\and E.~Merlin\orcid{0000-0001-6870-8900}\inst{\ref{aff25}}
\and G.~Meylan\orcid{0000-0001-6503-0209}\inst{\ref{aff2}}
\and P.~Monaco\orcid{0000-0003-2083-7564}\inst{\ref{aff76},\ref{aff12},\ref{aff13},\ref{aff11}}
\and A.~Mora\orcid{0000-0002-1922-8529}\inst{\ref{aff77}}
\and M.~Moresco\orcid{0000-0002-7616-7136}\inst{\ref{aff71},\ref{aff10}}
\and C.~Moretti\orcid{0000-0003-3314-8936}\inst{\ref{aff12},\ref{aff11},\ref{aff13}}
\and L.~Moscardini\orcid{0000-0002-3473-6716}\inst{\ref{aff71},\ref{aff10},\ref{aff16}}
\and R.~Nakajima\orcid{0009-0009-1213-7040}\inst{\ref{aff69}}
\and C.~Neissner\orcid{0000-0001-8524-4968}\inst{\ref{aff78},\ref{aff28}}
\and S.-M.~Niemi\orcid{0009-0005-0247-0086}\inst{\ref{aff79}}
\and J.~W.~Nightingale\orcid{0000-0002-8987-7401}\inst{\ref{aff80}}
\and C.~Padilla\orcid{0000-0001-7951-0166}\inst{\ref{aff78}}
\and K.~Paech\orcid{0000-0003-0625-2367}\inst{\ref{aff6}}
\and S.~Paltani\orcid{0000-0002-8108-9179}\inst{\ref{aff42}}
\and F.~Pasian\orcid{0000-0002-4869-3227}\inst{\ref{aff12}}
\and K.~Pedersen\inst{\ref{aff81}}
\and W.~J.~Percival\orcid{0000-0002-0644-5727}\inst{\ref{aff82},\ref{aff83},\ref{aff84}}
\and V.~Pettorino\orcid{0000-0002-4203-9320}\inst{\ref{aff79}}
\and A.~Pezzotta\orcid{0000-0003-0726-2268}\inst{\ref{aff9}}
\and S.~Pires\orcid{0000-0002-0249-2104}\inst{\ref{aff41}}
\and G.~Polenta\orcid{0000-0003-4067-9196}\inst{\ref{aff46}}
\and M.~Poncet\inst{\ref{aff85}}
\and L.~A.~Popa\inst{\ref{aff86}}
\and L.~Pozzetti\orcid{0000-0001-7085-0412}\inst{\ref{aff10}}
\and F.~Raison\orcid{0000-0002-7819-6918}\inst{\ref{aff6}}
\and R.~Rebolo\orcid{0000-0003-3767-7085}\inst{\ref{aff31},\ref{aff87},\ref{aff88}}
\and A.~Renzi\orcid{0000-0001-9856-1970}\inst{\ref{aff89},\ref{aff90},\ref{aff10}}
\and J.~Rhodes\orcid{0000-0002-4485-8549}\inst{\ref{aff55}}
\and G.~Riccio\inst{\ref{aff20}}
\and I.~Risso\orcid{0000-0003-2525-7761}\inst{\ref{aff9},\ref{aff18}}
\and E.~Romelli\orcid{0000-0003-3069-9222}\inst{\ref{aff12}}
\and M.~Roncarelli\orcid{0000-0001-9587-7822}\inst{\ref{aff10}}
\and B.~Rusholme\orcid{0000-0001-7648-4142}\inst{\ref{aff91}}
\and R.~Saglia\orcid{0000-0003-0378-7032}\inst{\ref{aff52},\ref{aff6}}
\and Z.~Sakr\orcid{0000-0002-4823-3757}\inst{\ref{aff92},\ref{aff93},\ref{aff94}}
\and D.~Sapone\orcid{0000-0001-7089-4503}\inst{\ref{aff95}}
\and B.~Sartoris\orcid{0000-0003-1337-5269}\inst{\ref{aff52},\ref{aff12}}
\and P.~Schneider\orcid{0000-0001-8561-2679}\inst{\ref{aff69}}
\and T.~Schrabback\orcid{0000-0002-6987-7834}\inst{\ref{aff96}}
\and A.~Secroun\orcid{0000-0003-0505-3710}\inst{\ref{aff44}}
\and E.~Sihvola\orcid{0000-0003-1804-7715}\inst{\ref{aff97}}
\and P.~Simon\inst{\ref{aff69}}
\and C.~Sirignano\orcid{0000-0002-0995-7146}\inst{\ref{aff89},\ref{aff90}}
\and G.~Sirri\orcid{0000-0003-2626-2853}\inst{\ref{aff16}}
\and J.~Skottfelt\orcid{0000-0003-1310-8283}\inst{\ref{aff98}}
\and L.~Stanco\orcid{0000-0002-9706-5104}\inst{\ref{aff90}}
\and P.~Tallada-Cresp\'{i}\orcid{0000-0002-1336-8328}\inst{\ref{aff27},\ref{aff28}}
\and A.~N.~Taylor\inst{\ref{aff32}}
\and I.~Tereno\orcid{0000-0002-4537-6218}\inst{\ref{aff99},\ref{aff100}}
\and N.~Tessore\orcid{0000-0002-9696-7931}\inst{\ref{aff40}}
\and S.~Toft\orcid{0000-0003-3631-7176}\inst{\ref{aff101},\ref{aff102}}
\and R.~Toledo-Moreo\orcid{0000-0002-2997-4859}\inst{\ref{aff103},\ref{aff104}}
\and F.~Torradeflot\orcid{0000-0003-1160-1517}\inst{\ref{aff28},\ref{aff27}}
\and I.~Tutusaus\orcid{0000-0002-3199-0399}\inst{\ref{aff4},\ref{aff47},\ref{aff93}}
\and J.~Valiviita\orcid{0000-0001-6225-3693}\inst{\ref{aff64},\ref{aff65}}
\and T.~Vassallo\orcid{0000-0001-6512-6358}\inst{\ref{aff12},\ref{aff50}}
\and G.~Verdoes~Kleijn\orcid{0000-0001-5803-2580}\inst{\ref{aff105}}
\and A.~Veropalumbo\orcid{0000-0003-2387-1194}\inst{\ref{aff9},\ref{aff18},\ref{aff17}}
\and Y.~Wang\orcid{0000-0002-4749-2984}\inst{\ref{aff91}}
\and J.~Weller\orcid{0000-0002-8282-2010}\inst{\ref{aff52},\ref{aff6}}
\and A.~Zacchei\orcid{0000-0003-0396-1192}\inst{\ref{aff12},\ref{aff11}}
\and G.~Zamorani\orcid{0000-0002-2318-301X}\inst{\ref{aff10}}
\and F.~M.~Zerbi\orcid{0000-0002-9996-973X}\inst{\ref{aff9}}
\and E.~Zucca\orcid{0000-0002-5845-8132}\inst{\ref{aff10}}
\and M.~Ballardini\orcid{0000-0003-4481-3559}\inst{\ref{aff106},\ref{aff107},\ref{aff10}}
\and E.~Bozzo\orcid{0000-0002-8201-1525}\inst{\ref{aff42}}
\and C.~Burigana\orcid{0000-0002-3005-5796}\inst{\ref{aff108},\ref{aff109}}
\and R.~Cabanac\orcid{0000-0001-6679-2600}\inst{\ref{aff93}}
\and A.~Cappi\inst{\ref{aff110},\ref{aff10}}
\and T.~Castro\orcid{0000-0002-6292-3228}\inst{\ref{aff12},\ref{aff13},\ref{aff11},\ref{aff111}}
\and J.~A.~Escartin~Vigo\inst{\ref{aff6}}
\and J.~Garc\'ia-Bellido\orcid{0000-0002-9370-8360}\inst{\ref{aff92}}
\and J.~Macias-Perez\orcid{0000-0002-5385-2763}\inst{\ref{aff112}}
\and R.~Maoli\orcid{0000-0002-6065-3025}\inst{\ref{aff113},\ref{aff25}}
\and J.~Mart\'{i}n-Fleitas\orcid{0000-0002-8594-569X}\inst{\ref{aff114}}
\and R.~B.~Metcalf\orcid{0000-0003-3167-2574}\inst{\ref{aff71},\ref{aff10}}
\and M.~P\"ontinen\orcid{0000-0001-5442-2530}\inst{\ref{aff64}}
\and V.~Scottez\orcid{0009-0008-3864-940X}\inst{\ref{aff115},\ref{aff116}}
\and M.~Sereno\orcid{0000-0003-0302-0325}\inst{\ref{aff10},\ref{aff16}}
\and M.~Tenti\orcid{0000-0002-4254-5901}\inst{\ref{aff16}}
\and M.~Tucci\inst{\ref{aff42}}
\and M.~Wiesmann\orcid{0009-0000-8199-5860}\inst{\ref{aff53}}
\and Y.~Akrami\orcid{0000-0002-2407-7956}\inst{\ref{aff92},\ref{aff117}}
\and I.~T.~Andika\orcid{0000-0001-6102-9526}\inst{\ref{aff52}}
\and G.~Angora\orcid{0000-0002-0316-6562}\inst{\ref{aff20},\ref{aff106}}
\and S.~Anselmi\orcid{0000-0002-3579-9583}\inst{\ref{aff90},\ref{aff89},\ref{aff118}}
\and M.~Archidiacono\orcid{0000-0003-4952-9012}\inst{\ref{aff119},\ref{aff120}}
\and F.~Atrio-Barandela\orcid{0000-0002-2130-2513}\inst{\ref{aff121}}
\and L.~Bazzanini\orcid{0000-0003-0727-0137}\inst{\ref{aff106},\ref{aff10}}
\and P.~Bergamini\orcid{0000-0003-1383-9414}\inst{\ref{aff10}}
\and D.~Bertacca\orcid{0000-0002-2490-7139}\inst{\ref{aff89},\ref{aff51},\ref{aff90}}
\and M.~Bethermin\orcid{0000-0002-3915-2015}\inst{\ref{aff5}}
\and F.~Beutler\orcid{0000-0003-0467-5438}\inst{\ref{aff32}}
\and L.~Blot\orcid{0000-0002-9622-7167}\inst{\ref{aff122},\ref{aff66}}
\and M.~Bonici\orcid{0000-0002-8430-126X}\inst{\ref{aff82},\ref{aff24}}
\and S.~Borgani\orcid{0000-0001-6151-6439}\inst{\ref{aff76},\ref{aff11},\ref{aff12},\ref{aff13},\ref{aff111}}
\and M.~L.~Brown\orcid{0000-0002-0370-8077}\inst{\ref{aff33}}
\and S.~Bruton\orcid{0000-0002-6503-5218}\inst{\ref{aff123}}
\and F.~Buitrago\orcid{0000-0002-2861-9812}\inst{\ref{aff124},\ref{aff125},\ref{aff100}}
\and A.~Calabro\orcid{0000-0003-2536-1614}\inst{\ref{aff25}}
\and B.~Camacho~Quevedo\orcid{0000-0002-8789-4232}\inst{\ref{aff11},\ref{aff14},\ref{aff12}}
\and F.~Caro\orcid{0009-0003-1053-0507}\inst{\ref{aff25}}
\and C.~S.~Carvalho\inst{\ref{aff100}}
\and Y.~Charles\orcid{0009-0000-3636-834X}\inst{\ref{aff70}}
\and F.~Cogato\orcid{0000-0003-4632-6113}\inst{\ref{aff71},\ref{aff10}}
\and S.~Conseil\orcid{0000-0002-3657-4191}\inst{\ref{aff35}}
\and A.~R.~Cooray\orcid{0000-0002-3892-0190}\inst{\ref{aff126}}
\and O.~Cucciati\orcid{0000-0002-9336-7551}\inst{\ref{aff10}}
\and S.~Davini\orcid{0000-0003-3269-1718}\inst{\ref{aff18}}
\and T.~de~Boer\orcid{0000-0001-5486-2747}\inst{\ref{aff127}}
\and F.~De~Paolis\orcid{0000-0001-6460-7563}\inst{\ref{aff128},\ref{aff129},\ref{aff130}}
\and G.~Desprez\orcid{0000-0001-8325-1742}\inst{\ref{aff105}}
\and A.~D\'iaz-S\'anchez\orcid{0000-0003-0748-4768}\inst{\ref{aff131}}
\and S.~Di~Domizio\orcid{0000-0003-2863-5895}\inst{\ref{aff17},\ref{aff18}}
\and J.~M.~Diego\orcid{0000-0001-9065-3926}\inst{\ref{aff132}}
\and V.~Duret\orcid{0009-0009-0383-4960}\inst{\ref{aff44}}
\and M.~Y.~Elkhashab\orcid{0000-0001-9306-2603}\inst{\ref{aff12},\ref{aff13},\ref{aff76},\ref{aff11}}
\and Y.~Fang\orcid{0000-0002-0334-6950}\inst{\ref{aff52}}
\and A.~Finoguenov\orcid{0000-0002-4606-5403}\inst{\ref{aff64}}
\and A.~Franco\orcid{0000-0002-4761-366X}\inst{\ref{aff129},\ref{aff128},\ref{aff130}}
\and K.~Ganga\orcid{0000-0001-8159-8208}\inst{\ref{aff73}}
\and T.~Gasparetto\orcid{0000-0002-7913-4866}\inst{\ref{aff25}}
\and R.~Gavazzi\orcid{0000-0002-5540-6935}\inst{\ref{aff70},\ref{aff133}}
\and E.~Gaztanaga\orcid{0000-0001-9632-0815}\inst{\ref{aff4},\ref{aff47},\ref{aff134}}
\and F.~Giacomini\orcid{0000-0002-3129-2814}\inst{\ref{aff16}}
\and F.~Gianotti\orcid{0000-0003-4666-119X}\inst{\ref{aff10}}
\and G.~Gozaliasl\orcid{0000-0002-0236-919X}\inst{\ref{aff135},\ref{aff64}}
\and A.~Gruppuso\orcid{0000-0001-9272-5292}\inst{\ref{aff10},\ref{aff16}}
\and M.~Guidi\orcid{0000-0001-9408-1101}\inst{\ref{aff15},\ref{aff10}}
\and C.~M.~Gutierrez\orcid{0000-0001-7854-783X}\inst{\ref{aff31},\ref{aff88}}
\and A.~Hall\orcid{0000-0002-3139-8651}\inst{\ref{aff32}}
\and C.~Hern\'andez-Monteagudo\orcid{0000-0001-5471-9166}\inst{\ref{aff88},\ref{aff31}}
\and H.~Hildebrandt\orcid{0000-0002-9814-3338}\inst{\ref{aff136}}
\and J.~Hjorth\orcid{0000-0002-4571-2306}\inst{\ref{aff81}}
\and J.~J.~E.~Kajava\orcid{0000-0002-3010-8333}\inst{\ref{aff137},\ref{aff138},\ref{aff139}}
\and Y.~Kang\orcid{0009-0000-8588-7250}\inst{\ref{aff42}}
\and V.~Kansal\orcid{0000-0002-4008-6078}\inst{\ref{aff140},\ref{aff141}}
\and D.~Karagiannis\orcid{0000-0002-4927-0816}\inst{\ref{aff106},\ref{aff142}}
\and K.~Kiiveri\orcid{0000-0002-3711-3346}\inst{\ref{aff97}}
\and J.~Kim\orcid{0000-0003-2776-2761}\inst{\ref{aff49}}
\and C.~C.~Kirkpatrick\inst{\ref{aff97}}
\and K.~Koyama\orcid{0000-0001-6727-6915}\inst{\ref{aff134}}
\and S.~Kruk\orcid{0000-0001-8010-8879}\inst{\ref{aff8}}
\and M.~C.~Lam\orcid{0000-0002-9347-2298}\inst{\ref{aff32}}
\and F.~Lepori\orcid{0009-0000-5061-7138}\inst{\ref{aff143}}
\and G.~Leroy\orcid{0009-0004-2523-4425}\inst{\ref{aff144},\ref{aff72}}
\and G.~F.~Lesci\orcid{0000-0002-4607-2830}\inst{\ref{aff71},\ref{aff10}}
\and J.~Lesgourgues\orcid{0000-0001-7627-353X}\inst{\ref{aff145}}
\and T.~I.~Liaudat\orcid{0000-0002-9104-314X}\inst{\ref{aff146}}
\and S.~J.~Liu\orcid{0000-0001-7680-2139}\inst{\ref{aff45}}
\and A.~Loureiro\orcid{0000-0002-4371-0876}\inst{\ref{aff147},\ref{aff148}}
\and G.~Maggio\orcid{0000-0003-4020-4836}\inst{\ref{aff12}}
\and F.~Mannucci\orcid{0000-0002-4803-2381}\inst{\ref{aff149}}
\and C.~J.~A.~P.~Martins\orcid{0000-0002-4886-9261}\inst{\ref{aff150},\ref{aff151}}
\and M.~Miluzio\inst{\ref{aff8},\ref{aff152}}
\and G.~Morgante\inst{\ref{aff10}}
\and C.~Murray\inst{\ref{aff73}}
\and S.~Nadathur\orcid{0000-0001-9070-3102}\inst{\ref{aff134}}
\and K.~Naidoo\orcid{0000-0002-9182-1802}\inst{\ref{aff134},\ref{aff60}}
\and A.~Navarro-Alsina\orcid{0000-0002-3173-2592}\inst{\ref{aff69}}
\and S.~Nesseris\orcid{0000-0002-0567-0324}\inst{\ref{aff92}}
\and F.~Pace\orcid{0000-0001-8039-0480}\inst{\ref{aff21},\ref{aff22},\ref{aff23}}
\and D.~Paoletti\orcid{0000-0003-4761-6147}\inst{\ref{aff10},\ref{aff109}}
\and F.~Passalacqua\orcid{0000-0002-8606-4093}\inst{\ref{aff89},\ref{aff90}}
\and K.~Paterson\orcid{0000-0001-8340-3486}\inst{\ref{aff60}}
\and L.~Patrizii\inst{\ref{aff16}}
\and C.~Pattison\orcid{0000-0003-3272-2617}\inst{\ref{aff134}}
\and A.~Pisani\orcid{0000-0002-6146-4437}\inst{\ref{aff44}}
\and D.~Potter\orcid{0000-0002-0757-5195}\inst{\ref{aff153}}
\and G.~W.~Pratt\inst{\ref{aff41}}
\and S.~Quai\orcid{0000-0002-0449-8163}\inst{\ref{aff71},\ref{aff10}}
\and M.~Radovich\orcid{0000-0002-3585-866X}\inst{\ref{aff51}}
\and K.~Rojas\orcid{0000-0003-1391-6854}\inst{\ref{aff154}}
\and W.~Roster\orcid{0000-0002-9149-6528}\inst{\ref{aff6}}
\and S.~Sacquegna\orcid{0000-0002-8433-6630}\inst{\ref{aff155}}
\and M.~Sahl\'en\orcid{0000-0003-0973-4804}\inst{\ref{aff156}}
\and D.~B.~Sanders\orcid{0000-0002-1233-9998}\inst{\ref{aff127}}
\and E.~Sarpa\orcid{0000-0002-1256-655X}\inst{\ref{aff12}}
\and A.~Schneider\orcid{0000-0001-7055-8104}\inst{\ref{aff153}}
\and M.~Schultheis\inst{\ref{aff110}}
\and D.~Sciotti\orcid{0009-0008-4519-2620}\inst{\ref{aff25},\ref{aff26}}
\and E.~Sellentin\inst{\ref{aff157},\ref{aff54}}
\and F.~Shankar\orcid{0000-0001-8973-5051}\inst{\ref{aff158}}
\and L.~C.~Smith\orcid{0000-0002-3259-2771}\inst{\ref{aff159}}
\and J.~Stadel\orcid{0000-0001-7565-8622}\inst{\ref{aff153}}
\and I.~Szapudi\orcid{0000-0003-2274-0301}\inst{\ref{aff127}}
\and K.~Tanidis\orcid{0000-0001-9843-5130}\inst{\ref{aff160}}
\and F.~Tarsitano\orcid{0000-0002-5919-0238}\inst{\ref{aff161},\ref{aff42}}
\and G.~Testera\orcid{0000-0003-2970-766X}\inst{\ref{aff18}}
\and R.~Teyssier\orcid{0000-0001-7689-0933}\inst{\ref{aff162}}
\and S.~Tosi\orcid{0000-0002-7275-9193}\inst{\ref{aff17},\ref{aff9},\ref{aff18}}
\and A.~Troja\orcid{0000-0003-0239-4595}\inst{\ref{aff12}}
\and C.~Uhlemann\orcid{0000-0001-7831-1579}\inst{\ref{aff163},\ref{aff80}}
\and C.~Valieri\inst{\ref{aff16}}
\and A.~Venhola\orcid{0000-0001-6071-4564}\inst{\ref{aff164}}
\and D.~Vergani\orcid{0000-0003-0898-2216}\inst{\ref{aff10}}
\and G.~Verza\orcid{0000-0002-1886-8348}\thanks{Deceased}\inst{\ref{aff165},\ref{aff166}}
\and E.~Vilenius\orcid{0000-0002-6184-7681}\inst{\ref{aff40}}
\and S.~Vinciguerra\orcid{0009-0005-4018-3184}\inst{\ref{aff70}}
\and M.~von~Wietersheim-Kramsta\orcid{0000-0003-4986-5091}\inst{\ref{aff72},\ref{aff144}}
\and N.~A.~Walton\orcid{0000-0003-3983-8778}\inst{\ref{aff159}}
\and J.~R.~Weaver\orcid{0000-0003-1614-196X}\inst{\ref{aff167}}
\and A.~H.~Wright\orcid{0000-0001-7363-7932}\inst{\ref{aff136}}
\and H.~W.~Yeung\orcid{0000-0002-4993-9014}\inst{\ref{aff32}}
\and J.~H.~Knapen\orcid{0000-0003-1643-0024}\inst{\ref{aff31},\ref{aff88}}}
										   
\institute{School of Physics and Astronomy, University of Nottingham, University Park, Nottingham NG7 2RD, UK\label{aff1}
\and
Institute of Physics, Laboratory of Astrophysics, Ecole Polytechnique F\'ed\'erale de Lausanne (EPFL), Observatoire de Sauverny, 1290 Versoix, Switzerland\label{aff2}
\and
Department of Astronomy, University of Florida, Bryant Space Science Center, Gainesville, FL 32611, USA\label{aff3}
\and
Institute of Space Sciences (ICE, CSIC), Campus UAB, Carrer de Can Magrans, s/n, 08193 Barcelona, Spain\label{aff4}
\and
Universit\'e de Strasbourg, CNRS, Observatoire astronomique de Strasbourg, UMR 7550, 67000 Strasbourg, France\label{aff5}
\and
Max Planck Institute for Extraterrestrial Physics, Giessenbachstr. 1, 85748 Garching, Germany\label{aff6}
\and
Department of Physics and Astronomy, University of British Columbia, Vancouver, BC V6T 1Z1, Canada\label{aff7}
\and
ESAC/ESA, Camino Bajo del Castillo, s/n., Urb. Villafranca del Castillo, 28692 Villanueva de la Ca\~nada, Madrid, Spain\label{aff8}
\and
INAF-Osservatorio Astronomico di Brera, Via Brera 28, 20122 Milano, Italy\label{aff9}
\and
INAF-Osservatorio di Astrofisica e Scienza dello Spazio di Bologna, Via Piero Gobetti 93/3, 40129 Bologna, Italy\label{aff10}
\and
IFPU, Institute for Fundamental Physics of the Universe, via Beirut 2, 34151 Trieste, Italy\label{aff11}
\and
INAF-Osservatorio Astronomico di Trieste, Via G. B. Tiepolo 11, 34143 Trieste, Italy\label{aff12}
\and
INFN, Sezione di Trieste, Via Valerio 2, 34127 Trieste TS, Italy\label{aff13}
\and
SISSA, International School for Advanced Studies, Via Bonomea 265, 34136 Trieste TS, Italy\label{aff14}
\and
Dipartimento di Fisica e Astronomia, Universit\`a di Bologna, Via Gobetti 93/2, 40129 Bologna, Italy\label{aff15}
\and
INFN-Sezione di Bologna, Viale Berti Pichat 6/2, 40127 Bologna, Italy\label{aff16}
\and
Dipartimento di Fisica, Universit\`a di Genova, Via Dodecaneso 33, 16146, Genova, Italy\label{aff17}
\and
INFN-Sezione di Genova, Via Dodecaneso 33, 16146, Genova, Italy\label{aff18}
\and
Department of Physics "E. Pancini", University Federico II, Via Cinthia 6, 80126, Napoli, Italy\label{aff19}
\and
INAF-Osservatorio Astronomico di Capodimonte, Via Moiariello 16, 80131 Napoli, Italy\label{aff20}
\and
Dipartimento di Fisica, Universit\`a degli Studi di Torino, Via P. Giuria 1, 10125 Torino, Italy\label{aff21}
\and
INFN-Sezione di Torino, Via P. Giuria 1, 10125 Torino, Italy\label{aff22}
\and
INAF-Osservatorio Astrofisico di Torino, Via Osservatorio 20, 10025 Pino Torinese (TO), Italy\label{aff23}
\and
INAF-IASF Milano, Via Alfonso Corti 12, 20133 Milano, Italy\label{aff24}
\and
INAF-Osservatorio Astronomico di Roma, Via Frascati 33, 00078 Monteporzio Catone, Italy\label{aff25}
\and
INFN-Sezione di Roma, Piazzale Aldo Moro, 2 - c/o Dipartimento di Fisica, Edificio G. Marconi, 00185 Roma, Italy\label{aff26}
\and
Centro de Investigaciones Energ\'eticas, Medioambientales y Tecnol\'ogicas (CIEMAT), Avenida Complutense 40, 28040 Madrid, Spain\label{aff27}
\and
Port d'Informaci\'{o} Cient\'{i}fica, Campus UAB, C. Albareda s/n, 08193 Bellaterra (Barcelona), Spain\label{aff28}
\and
INFN -- Sezione di Napoli, Via Cinthia 6, 80126, Napoli, Italy\label{aff29}
\and
Dipartimento di Fisica e Astronomia "Augusto Righi" - Alma Mater Studiorum Universit\`a di Bologna, Viale Berti Pichat 6/2, 40127 Bologna, Italy\label{aff30}
\and
Instituto de Astrof\'{\i}sica de Canarias, E-38205 La Laguna, Tenerife, Spain\label{aff31}
\and
Institute for Astronomy, University of Edinburgh, Royal Observatory, Blackford Hill, Edinburgh EH9 3HJ, UK\label{aff32}
\and
Jodrell Bank Centre for Astrophysics, Department of Physics and Astronomy, University of Manchester, Oxford Road, Manchester M13 9PL, UK\label{aff33}
\and
European Space Agency/ESRIN, Largo Galileo Galilei 1, 00044 Frascati, Roma, Italy\label{aff34}
\and
Universit\'e Claude Bernard Lyon 1, CNRS/IN2P3, IP2I Lyon, UMR 5822, Villeurbanne, F-69100, France\label{aff35}
\and
Institut de Ci\`{e}ncies del Cosmos (ICCUB), Universitat de Barcelona (IEEC-UB), Mart\'{i} i Franqu\`{e}s 1, 08028 Barcelona, Spain\label{aff36}
\and
Instituci\'o Catalana de Recerca i Estudis Avan\c{c}ats (ICREA), Passeig de Llu\'{\i}s Companys 23, 08010 Barcelona, Spain\label{aff37}
\and
Institut de Ciencies de l'Espai (IEEC-CSIC), Campus UAB, Carrer de Can Magrans, s/n Cerdanyola del Vall\'es, 08193 Barcelona, Spain\label{aff38}
\and
UCB Lyon 1, CNRS/IN2P3, IUF, IP2I Lyon, 4 rue Enrico Fermi, 69622 Villeurbanne, France\label{aff39}
\and
Mullard Space Science Laboratory, University College London, Holmbury St Mary, Dorking, Surrey RH5 6NT, UK\label{aff40}
\and
Universit\'e Paris-Saclay, Universit\'e Paris Cit\'e, CEA, CNRS, AIM, 91191, Gif-sur-Yvette, France\label{aff41}
\and
Department of Astronomy, University of Geneva, ch. d'Ecogia 16, 1290 Versoix, Switzerland\label{aff42}
\and
Universit\'e Paris-Saclay, CNRS, Institut d'astrophysique spatiale, 91405, Orsay, France\label{aff43}
\and
Aix-Marseille Universit\'e, CNRS/IN2P3, CPPM, Marseille, France\label{aff44}
\and
INAF-Istituto di Astrofisica e Planetologia Spaziali, via del Fosso del Cavaliere, 100, 00100 Roma, Italy\label{aff45}
\and
Space Science Data Center, Italian Space Agency, via del Politecnico snc, 00133 Roma, Italy\label{aff46}
\and
Institut d'Estudis Espacials de Catalunya (IEEC),  Edifici RDIT, Campus UPC, 08860 Castelldefels, Barcelona, Spain\label{aff47}
\and
School of Physics, HH Wills Physics Laboratory, University of Bristol, Tyndall Avenue, Bristol, BS8 1TL, UK\label{aff48}
\and
Department of Physics, University of Oxford, Keble Road, Oxford OX1 3RH, UK\label{aff49}
\and
University Observatory, LMU Faculty of Physics, Scheinerstr.~1, 81679 Munich, Germany\label{aff50}
\and
INAF-Osservatorio Astronomico di Padova, Via dell'Osservatorio 5, 35122 Padova, Italy\label{aff51}
\and
Universit\"ats-Sternwarte M\"unchen, Fakult\"at f\"ur Physik, Ludwig-Maximilians-Universit\"at M\"unchen, Scheinerstr.~1, 81679 M\"unchen, Germany\label{aff52}
\and
Institute of Theoretical Astrophysics, University of Oslo, P.O. Box 1029 Blindern, 0315 Oslo, Norway\label{aff53}
\and
Leiden Observatory, Leiden University, Einsteinweg 55, 2333 CC Leiden, The Netherlands\label{aff54}
\and
Jet Propulsion Laboratory, California Institute of Technology, 4800 Oak Grove Drive, Pasadena, CA, 91109, USA\label{aff55}
\and
Department of Physics, Lancaster University, Lancaster, LA1 4YB, UK\label{aff56}
\and
Felix Hormuth Engineering, Goethestr. 17, 69181 Leimen, Germany\label{aff57}
\and
Technical University of Denmark, Elektrovej 327, 2800 Kgs. Lyngby, Denmark\label{aff58}
\and
Cosmic Dawn Center (DAWN), Denmark\label{aff59}
\and
Max-Planck-Institut f\"ur Astronomie, K\"onigstuhl 17, 69117 Heidelberg, Germany\label{aff60}
\and
NASA Goddard Space Flight Center, Greenbelt, MD 20771, USA\label{aff61}
\and
Department of Physics and Astronomy, University College London, Gower Street, London WC1E 6BT, UK\label{aff62}
\and
Universit\'e de Gen\`eve, D\'epartement de Physique Th\'eorique and Centre for Astroparticle Physics, 24 quai Ernest-Ansermet, CH-1211 Gen\`eve 4, Switzerland\label{aff63}
\and
Department of Physics, P.O. Box 64, University of Helsinki, 00014 Helsinki, Finland\label{aff64}
\and
Helsinki Institute of Physics, Gustaf H{\"a}llstr{\"o}min katu 2, University of Helsinki, 00014 Helsinki, Finland\label{aff65}
\and
Laboratoire d'etude de l'Univers et des phenomenes eXtremes, Observatoire de Paris, Universit\'e PSL, Sorbonne Universit\'e, CNRS, 92190 Meudon, France\label{aff66}
\and
SKAO, Jodrell Bank, Lower Withington, Macclesfield SK11 9FT, UK\label{aff67}
\and
Centre de Calcul de l'IN2P3/CNRS, 21 avenue Pierre de Coubertin 69627 Villeurbanne Cedex, France\label{aff68}
\and
Universit\"at Bonn, Argelander-Institut f\"ur Astronomie, Auf dem H\"ugel 71, 53121 Bonn, Germany\label{aff69}
\and
Aix-Marseille Universit\'e, CNRS, CNES, LAM, Marseille, France\label{aff70}
\and
Dipartimento di Fisica e Astronomia "Augusto Righi" - Alma Mater Studiorum Universit\`a di Bologna, via Piero Gobetti 93/2, 40129 Bologna, Italy\label{aff71}
\and
Department of Physics, Institute for Computational Cosmology, Durham University, South Road, Durham, DH1 3LE, UK\label{aff72}
\and
Universit\'e Paris Cit\'e, CNRS, Astroparticule et Cosmologie, 75013 Paris, France\label{aff73}
\and
CNRS-UCB International Research Laboratory, Centre Pierre Bin\'etruy, IRL2007, CPB-IN2P3, Berkeley, USA\label{aff74}
\and
University of Applied Sciences and Arts of Northwestern Switzerland, School of Engineering, 5210 Windisch, Switzerland\label{aff75}
\and
Dipartimento di Fisica - Sezione di Astronomia, Universit\`a di Trieste, Via Tiepolo 11, 34131 Trieste, Italy\label{aff76}
\and
Telespazio UK S.L. for European Space Agency (ESA), Camino bajo del Castillo, s/n, Urbanizacion Villafranca del Castillo, Villanueva de la Ca\~nada, 28692 Madrid, Spain\label{aff77}
\and
Institut de F\'{i}sica d'Altes Energies (IFAE), The Barcelona Institute of Science and Technology, Campus UAB, 08193 Bellaterra (Barcelona), Spain\label{aff78}
\and
European Space Agency/ESTEC, Keplerlaan 1, 2201 AZ Noordwijk, The Netherlands\label{aff79}
\and
School of Mathematics, Statistics and Physics, Newcastle University, Herschel Building, Newcastle-upon-Tyne, NE1 7RU, UK\label{aff80}
\and
DARK, Niels Bohr Institute, University of Copenhagen, Jagtvej 155, 2200 Copenhagen, Denmark\label{aff81}
\and
Waterloo Centre for Astrophysics, University of Waterloo, Waterloo, Ontario N2L 3G1, Canada\label{aff82}
\and
Department of Physics and Astronomy, University of Waterloo, Waterloo, Ontario N2L 3G1, Canada\label{aff83}
\and
Perimeter Institute for Theoretical Physics, Waterloo, Ontario N2L 2Y5, Canada\label{aff84}
\and
Centre National d'Etudes Spatiales -- Centre spatial de Toulouse, 18 avenue Edouard Belin, 31401 Toulouse Cedex 9, France\label{aff85}
\and
Institute of Space Science, Str. Atomistilor, nr. 409 M\u{a}gurele, Ilfov, 077125, Romania\label{aff86}
\and
Consejo Superior de Investigaciones Cientificas, Calle Serrano 117, 28006 Madrid, Spain\label{aff87}
\and
Universidad de La Laguna, Dpto. Astrof\'\i sica, E-38206 La Laguna, Tenerife, Spain\label{aff88}
\and
Dipartimento di Fisica e Astronomia "G. Galilei", Universit\`a di Padova, Via Marzolo 8, 35131 Padova, Italy\label{aff89}
\and
INFN-Padova, Via Marzolo 8, 35131 Padova, Italy\label{aff90}
\and
Caltech/IPAC, 1200 E. California Blvd., Pasadena, CA 91125, USA\label{aff91}
\and
Instituto de F\'isica Te\'orica UAM-CSIC, Campus de Cantoblanco, 28049 Madrid, Spain\label{aff92}
\and
Institut de Recherche en Astrophysique et Plan\'etologie (IRAP), Universit\'e de Toulouse, CNRS, UPS, CNES, 14 Av. Edouard Belin, 31400 Toulouse, France\label{aff93}
\and
Universit\'e St Joseph; Faculty of Sciences, Beirut, Lebanon\label{aff94}
\and
Departamento de F\'isica, FCFM, Universidad de Chile, Blanco Encalada 2008, Santiago, Chile\label{aff95}
\and
Universit\"at Innsbruck, Institut f\"ur Astro- und Teilchenphysik, Technikerstr. 25/8, 6020 Innsbruck, Austria\label{aff96}
\and
Department of Physics and Helsinki Institute of Physics, Gustaf H\"allstr\"omin katu 2, University of Helsinki, 00014 Helsinki, Finland\label{aff97}
\and
Centre for Electronic Imaging, Open University, Walton Hall, Milton Keynes, MK7~6AA, UK\label{aff98}
\and
Departamento de F\'isica, Faculdade de Ci\^encias, Universidade de Lisboa, Edif\'icio C8, Campo Grande, PT1749-016 Lisboa, Portugal\label{aff99}
\and
Instituto de Astrof\'isica e Ci\^encias do Espa\c{c}o, Faculdade de Ci\^encias, Universidade de Lisboa, Tapada da Ajuda, 1349-018 Lisboa, Portugal\label{aff100}
\and
Cosmic Dawn Center (DAWN)\label{aff101}
\and
Niels Bohr Institute, University of Copenhagen, Jagtvej 128, 2200 Copenhagen, Denmark\label{aff102}
\and
Universidad Polit\'ecnica de Cartagena, Departamento de Electr\'onica y Tecnolog\'ia de Computadoras,  Plaza del Hospital 1, 30202 Cartagena, Spain\label{aff103}
\and
European University of Technology EUt+, European Union\label{aff104}
\and
Kapteyn Astronomical Institute, University of Groningen, PO Box 800, 9700 AV Groningen, The Netherlands\label{aff105}
\and
Dipartimento di Fisica e Scienze della Terra, Universit\`a degli Studi di Ferrara, Via Giuseppe Saragat 1, 44122 Ferrara, Italy\label{aff106}
\and
Istituto Nazionale di Fisica Nucleare, Sezione di Ferrara, Via Giuseppe Saragat 1, 44122 Ferrara, Italy\label{aff107}
\and
INAF, Istituto di Radioastronomia, Via Piero Gobetti 101, 40129 Bologna, Italy\label{aff108}
\and
INFN-Bologna, Via Irnerio 46, 40126 Bologna, Italy\label{aff109}
\and
Universit\'e C\^{o}te d'Azur, Observatoire de la C\^{o}te d'Azur, CNRS, Laboratoire Lagrange, Bd de l'Observatoire, CS 34229, 06304 Nice cedex 4, France\label{aff110}
\and
ICSC - Centro Nazionale di Ricerca in High Performance Computing, Big Data e Quantum Computing, Via Magnanelli 2, Bologna, Italy\label{aff111}
\and
Univ. Grenoble Alpes, CNRS, Grenoble INP, LPSC-IN2P3, 53, Avenue des Martyrs, 38000 Grenoble, France\label{aff112}
\and
Dipartimento di Fisica, Sapienza Universit\`a di Roma, Piazzale Aldo Moro 2, 00185 Roma, Italy\label{aff113}
\and
Aurora Technology for European Space Agency (ESA), Camino bajo del Castillo, s/n, Urbanizacion Villafranca del Castillo, Villanueva de la Ca\~nada, 28692 Madrid, Spain\label{aff114}
\and
Institut d'Astrophysique de Paris, 98bis Boulevard Arago, 75014, Paris, France\label{aff115}
\and
ICL, Junia, Universit\'e Catholique de Lille, LITL, 59000 Lille, France\label{aff116}
\and
CERCA/ISO, Department of Physics, Case Western Reserve University, 10900 Euclid Avenue, Cleveland, OH 44106, USA\label{aff117}
\and
Laboratoire Univers et Th\'eorie, Observatoire de Paris, Universit\'e PSL, Universit\'e Paris Cit\'e, CNRS, 92190 Meudon, France\label{aff118}
\and
Dipartimento di Fisica "Aldo Pontremoli", Universit\`a degli Studi di Milano, Via Celoria 16, 20133 Milano, Italy\label{aff119}
\and
INFN-Sezione di Milano, Via Celoria 16, 20133 Milano, Italy\label{aff120}
\and
Departamento de F{\'\i}sica Fundamental. Universidad de Salamanca. Plaza de la Merced s/n. 37008 Salamanca, Spain\label{aff121}
\and
Center for Data-Driven Discovery, Kavli IPMU (WPI), UTIAS, The University of Tokyo, Kashiwa, Chiba 277-8583, Japan\label{aff122}
\and
California Institute of Technology, 1200 E California Blvd, Pasadena, CA 91125, USA\label{aff123}
\and
Departamento de F\'{i}sica Te\'{o}rica, At\'{o}mica y \'{O}ptica, Universidad de Valladolid, 47011 Valladolid, Spain\label{aff124}
\and
Laboratory for Disruptive Interdisciplinary Science (LaDIS), Universidad de Valladolid, 47011 Valladolid, Spain\label{aff125}
\and
Department of Physics \& Astronomy, University of California Irvine, Irvine CA 92697, USA\label{aff126}
\and
Institute for Astronomy, University of Hawaii, 2680 Woodlawn Drive, Honolulu, HI 96822, USA\label{aff127}
\and
Department of Mathematics and Physics E. De Giorgi, University of Salento, Via per Arnesano, CP-I93, 73100, Lecce, Italy\label{aff128}
\and
INFN, Sezione di Lecce, Via per Arnesano, CP-193, 73100, Lecce, Italy\label{aff129}
\and
INAF-Sezione di Lecce, c/o Dipartimento Matematica e Fisica, Via per Arnesano, 73100, Lecce, Italy\label{aff130}
\and
Departamento F\'isica Aplicada, Universidad Polit\'ecnica de Cartagena, Campus Muralla del Mar, 30202 Cartagena, Murcia, Spain\label{aff131}
\and
Instituto de F\'isica de Cantabria, Edificio Juan Jord\'a, Avenida de los Castros, 39005 Santander, Spain\label{aff132}
\and
Institut d'Astrophysique de Paris, UMR 7095, CNRS, and Sorbonne Universit\'e, 98 bis boulevard Arago, 75014 Paris, France\label{aff133}
\and
Institute of Cosmology and Gravitation, University of Portsmouth, Portsmouth PO1 3FX, UK\label{aff134}
\and
Department of Computer Science, Aalto University, PO Box 15400, Espoo, FI-00 076, Finland\label{aff135}
\and
Ruhr University Bochum, Faculty of Physics and Astronomy, Astronomical Institute (AIRUB), German Centre for Cosmological Lensing (GCCL), 44780 Bochum, Germany\label{aff136}
\and
Department of Physics and Astronomy, Vesilinnantie 5, University of Turku, 20014 Turku, Finland\label{aff137}
\and
Finnish Centre for Astronomy with ESO (FINCA), Quantum, Vesilinnantie 5, University of Turku, 20014 Turku, Finland\label{aff138}
\and
Serco for European Space Agency (ESA), Camino bajo del Castillo, s/n, Urbanizacion Villafranca del Castillo, Villanueva de la Ca\~nada, 28692 Madrid, Spain\label{aff139}
\and
ARC Centre of Excellence for Dark Matter Particle Physics, Melbourne, Australia\label{aff140}
\and
Centre for Astrophysics \& Supercomputing, Swinburne University of Technology,  Hawthorn, Victoria 3122, Australia\label{aff141}
\and
Department of Physics and Astronomy, University of the Western Cape, Bellville, Cape Town, 7535, South Africa\label{aff142}
\and
Departement of Theoretical Physics, University of Geneva, Switzerland\label{aff143}
\and
Department of Physics, Centre for Extragalactic Astronomy, Durham University, South Road, Durham, DH1 3LE, UK\label{aff144}
\and
Institute for Theoretical Particle Physics and Cosmology (TTK), RWTH Aachen University, 52056 Aachen, Germany\label{aff145}
\and
IRFU, CEA, Universit\'e Paris-Saclay 91191 Gif-sur-Yvette Cedex, France\label{aff146}
\and
Oskar Klein Centre for Cosmoparticle Physics, Department of Physics, Stockholm University, Stockholm, SE-106 91, Sweden\label{aff147}
\and
Astrophysics Group, Blackett Laboratory, Imperial College London, London SW7 2AZ, UK\label{aff148}
\and
INAF-Osservatorio Astrofisico di Arcetri, Largo E. Fermi 5, 50125, Firenze, Italy\label{aff149}
\and
Centro de Astrof\'{\i}sica da Universidade do Porto, Rua das Estrelas, 4150-762 Porto, Portugal\label{aff150}
\and
Instituto de Astrof\'isica e Ci\^encias do Espa\c{c}o, Universidade do Porto, CAUP, Rua das Estrelas, PT4150-762 Porto, Portugal\label{aff151}
\and
HE Space for European Space Agency (ESA), Camino bajo del Castillo, s/n, Urbanizacion Villafranca del Castillo, Villanueva de la Ca\~nada, 28692 Madrid, Spain\label{aff152}
\and
Department of Astrophysics, University of Zurich, Winterthurerstrasse 190, 8057 Zurich, Switzerland\label{aff153}
\and
University of Applied Sciences and Arts of Northwestern Switzerland, School of Computer Science, 5210 Windisch, Switzerland\label{aff154}
\and
INAF - Osservatorio Astronomico d'Abruzzo, Via Maggini, 64100, Teramo, Italy\label{aff155}
\and
Theoretical astrophysics, Department of Physics and Astronomy, Uppsala University, Box 516, 751 37 Uppsala, Sweden\label{aff156}
\and
Mathematical Institute, University of Leiden, Einsteinweg 55, 2333 CA Leiden, The Netherlands\label{aff157}
\and
School of Physics \& Astronomy, University of Southampton, Highfield Campus, Southampton SO17 1BJ, UK\label{aff158}
\and
Institute of Astronomy, University of Cambridge, Madingley Road, Cambridge CB3 0HA, UK\label{aff159}
\and
Center for Astrophysics and Cosmology, University of Nova Gorica, Nova Gorica, Slovenia\label{aff160}
\and
Institute for Particle Physics and Astrophysics, Dept. of Physics, ETH Zurich, Wolfgang-Pauli-Strasse 27, 8093 Zurich, Switzerland\label{aff161}
\and
Department of Astrophysical Sciences, Peyton Hall, Princeton University, Princeton, NJ 08544, USA\label{aff162}
\and
Fakult\"at f\"ur Physik, Universit\"at Bielefeld, Postfach 100131, 33501 Bielefeld, Germany\label{aff163}
\and
Space physics and astronomy research unit, University of Oulu, Pentti Kaiteran katu 1, FI-90014 Oulu, Finland\label{aff164}
\and
International Centre for Theoretical Physics (ICTP), Strada Costiera 11, 34151 Trieste, Italy\label{aff165}
\and
Center for Computational Astrophysics, Flatiron Institute, 162 5th Avenue, 10010, New York, NY, USA\label{aff166}
\and
MIT Kavli Institute for Astrophysics and Space Research, Massachusetts Institute of Technology, Cambridge, MA 02139, USA\label{aff167}} 

\date{Received 27 July 2026 / Revised 9 October 2026}

 \abstract{
The \Euclid survey provides images with extraordinary potential for studies of the low-surface-brightness Universe, including intracluster light (ICL). However, to realise this promise, the data must be carefully treated to minimise background variations while preserving the signal from diffuse structures. In this work, we describe the post-processing and measurement procedures used for studies of ICL with \Euclid. We demonstrate that we are able to substantially reduce background variations without resorting to a small-scale background subtraction. By injecting simulated clusters into the real images, we show that our approach preserves the faint outskirts of extended sources. On scales of several arcminutes, we can reliably detect surface-brightness levels around \qty{27.5}{mag.arcsec^{-2}} in \YE, \JE, \HE, and \qty{29}{mag.arcsec^{-2}} in \IE. As a result, we can obtain accurate measurements of the ICL in individual clusters from the local Universe out to \changed{$z\sim 0.8$} with uncertainties on the order of 10\% in surface brightness, \ang{10} in position angle, and $0.1$ in ellipticity. Our ability to measure the diffuse light with \Euclid is thus set to meet expectations and revolutionise our understanding of the ICL.
}

\keywords{Methods: data analysis -- Techniques: image processing -- Galaxies: clusters: general -- Galaxies: halos}

   \titlerunning{Data processing for ICL}
   \authorrunning{Euclid Collaboration: S.~P. Bamford et al.}

   \maketitle

\section{\label{sc:Intro}Introduction}

\begin{figure*}[tbp]
\centering
\includegraphics[width=\hsize]{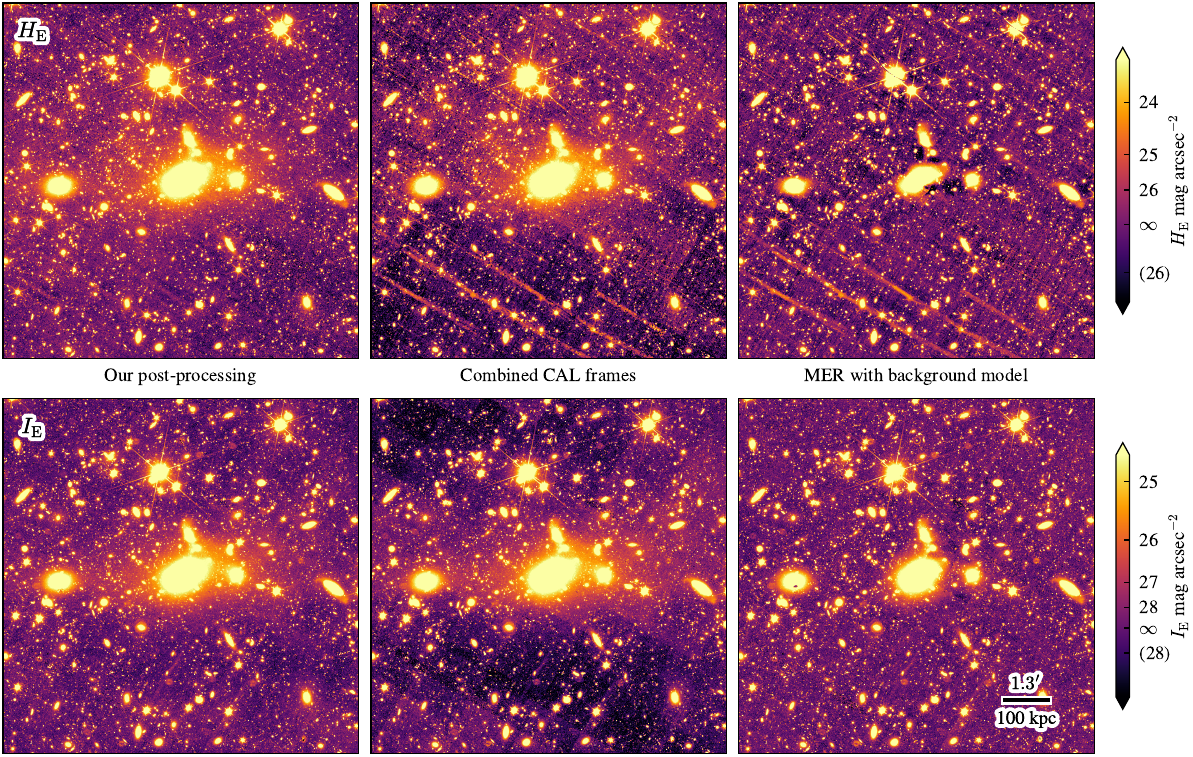}
\caption{\label{fig:nicl_vs_MER}Comparison of our ICL-focussed data processing (\emph{left}) versus LE2 calibrated frames combined with no post-processing (\emph{centre}) and the standard \Euclid MER product with the MER background model restored (\emph{right}), for \HE (\emph{top} row) and \IE (\emph{bottom} row). All images show an $11\arcmin \times 11\arcmin$ ($\qty{1}{Mpc} \times \qty{1}{Mpc}$) region of observation 2683, centred on the cluster MCXC~J1754.6+6803 at $z=0.077$ with $M_{200} \sim \num{1.4e14}\Msun$ and $R_{200} \sim \qty{0.8}{Mpc}$ \citep{2011A&A...534A.109P}. The images are displayed with an asinh flux scaling, surface brightness is indicated by the colourbars. For negative flux values, the equivalent surface brightness of a source with that absolute flux level is given in parentheses. The small-scale background subtraction applied by MER (\emph{right}) has mostly removed the extended diffuse emission that is clearly seen in the other two images. However, simply avoiding the background subtraction (\emph{centre}) leaves behind artefacts that would compromise the sensitivity of our LSB measurements. Our post-processing (\emph{left}) mostly removes these artefacts, while preserving the extended diffuse light.}
\end{figure*}

Tidal interactions between galaxies, or between galaxies and larger structures, result in distinctive morphological features. In particular, stars may be stripped from the main body of a galaxy to form tails, streams, and, ultimately, diffuse haloes (e.g., \citealt{1998Natur.396..549G,2005ApJ...635..931B,2015MNRAS.446..120D}).
The prevalence and characteristics of these features depend, among other things, on the radial profile of the gravitational potential of each interacting system and the distribution of stellar populations within each potential (e.g., \citealt{2012ApJS..202....8S,2024ApJ...966..124K,2025MNRAS.543.3391K}).
For example, the diffuse stellar halo of galaxy clusters, known as intracluster light (ICL; \citealt{2021Galax...9...60C,2022NatAs...6..308M}), can provide insights into populations of galaxies that have been disrupted in these dense environments, as well as revealing the shape and orientation of the dark matter-dominated cluster potential
(e.g., \citealt{2025A&A...703A..85C,2025MNRAS.539.2279B,2026arXiv260118693M}).
Since all these features result from the outskirts of galaxies being redistributed over a wider area, they typically have very low surface brightness (LSB).

The LSB Universe (see, for instance, \citealt{2017ASSL..434..255K} and references therein) thus contains a great deal of information valuable to our understanding of both galaxy evolution and the nature of dark matter.
To access this realm, we must overcome a series of technical challenges.
At best, the available information is limited only by the Poisson noise on the photon flux from the source itself.
In practice, however, other phenomena dominate the noise for all but the brightest sources.
Emission from the atmosphere limits the surface-brightness levels detectable in ground-based observations, but can be avoided by going to space.
Even there, LSB performance depends on how well one is able to avoid sources of background light and instrumental noise.
These cannot be entirely eliminated, so the surface-brightness limit is set by our ability to model and remove contamination, while avoiding damage to the target signal.
The most successful approaches therefore focus on simultaneously avoiding sources of background noise, while ensuring that what remains is smoothly distributed, and thus amenable to modelling.
A range of instrumentation design innovations \citep[e.g.,][]{1992SPIE.1656..414L,2003SPIE.4841...72B,2004SPIE.5492..484K,2008SPIE.7014E..0ED,2018PASJ...70S...1M} and data processing techniques \citep[e.g.,][]{2002ApJ...575..779F,2005ApJ...631L..41M, 2014PASP..126...55A,2015MNRAS.446..120D,2016ApJ...823..123T,2019A&A...621A.133B,2020ApJS..247...43K,2023MNRAS.520.2484K} have been previously developed for LSB observations.

The \Euclid mission \citep{EuclidSkyOverview} has a unique combination of a \qty{1}{m^2} collecting area and \qty{0.5}{deg^2} field of view (FoV) on a space-based platform, allowing it to efficiently survey the sky with an unprecedented combination of resolution (0.15\,arcsec in the visible and 0.45\,arcsec in the near-infrared) and depth.
The telescope as a whole \citep{Gaspar-Venancio23}, as well as its VIS \citep{EuclidSkyVIS} and NISP \citep{EuclidSkyNISP} instruments, have been carefully engineered to produce a compact and stable point spread function, and reduce various other sources of stray light.
Over six years, the Euclid Wide Survey (EWS; \citealt{Scaramella-EP1}) will image approximately \qty{14000}{deg^2} in the optical (obtained with VIS, at \qty{0.1}{arcsec.pixel^{-1}}) and near-infrared (NIR images, obtained with NISP, at \qty{0.3}{arcsec.pixel^{-1}}).
Several regions, totalling about \qty{60}{deg^2}, will be observed multiple times and hence reach substantially greater depth.
Together, these will provide a revolutionary data set for statistical studies of LSB features \citep{Borlaff-EP16}, utilising large numbers of objects distributed across the sky.
As an illustration, we expect to confidently detect ICL around \num{80000} galaxy clusters at $z=0.3$--$1.5$ in the full EWS \citep{EP-Bellhouse}.

The primary aim of the \Euclid mission is to constrain cosmological parameters through a combination of weak lensing and galaxy clustering measurements.
The desired performance of these cosmological probes sets the technical requirements of the mission and the priorities of the data processing teams.
Preserving large-scale LSB features was therefore not a primary concern in the development of the data reduction pipelines.
The processing is thus able to deal with a variety of issues by estimating the background on small angular scales and subtracting this from the images. Such a background subtraction is actually performed twice: first on the individual dithered exposures prior to combining, then again on the combined tiles. The first background is estimated in the VIS \citep{Q1-TP002} and NIR \citep{Q1-TP003} processing of individual observations, but both types of background subtraction are applied during the mosaicking process (MER; \citealt{Q1-TP004}).
The quality of the \Euclid VIS and NIR images processed by the Science Ground Segment (SGS) team is outstanding, meeting the technical requirements and setting the stage for delivering the cosmological science goals of the mission.

The small-scale background subtraction means that photometry of sources smaller than 10\,arcsec is highly accurate in the final MER tiles and catalogues. However, significant flux is removed from larger objects, as evident in \reffig{nicl_vs_MER}. The second background -- that estimated and subtracted from the combined images by MER -- is provided as a standard product and may be restored to the images. For VIS, this is sufficient for studying LSB sources up to one arcminute in size (such as tidal features around fairly distant interacting galaxies, see Sola et al. in prep., and dwarf galaxies at distances beyond 10\,Mpc, see \citealt{EP-Urbano,Q1-SP001}). However, for Q1 MER images in the NIR bands, the problem is more severe. Since the background subtracted from the individual dithers cannot be restored to the tiles, the Q1 MER products do not permit VIS+NIR studies of LSB sources at scales above 0.5\,arcmin (see \reffig{nicl_vs_MER} and \refsec{sb_limit}).
This is a common issue, with galaxy surveys often requiring specific processing for large and/or LSB sources (e.g., \citealt{2011AJ....142...31B,2019ApJ...874..165Z,2019A&A...621A.133B}). For ICL studies with \Euclid, some degree of alternative data processing is needed as well.

The potential of \Euclid data for ICL studies\footnote{Other LSB science cases, such as dwarf galaxies \citep{EROPerseusDGs}, extended haloes \citep{Mondelin25}, and tidal features \citep{Urbano24}, have also been illustrated using the same data.} has been spectacularly demonstrated by the Early Release Observations (ERO) of the Perseus Cluster \citep{EROPerseusICL} and Abell 2390 \citep{Ellien25}. These studies made use of a dedicated reduction of the ERO data \citep[][hereafter C25]{EROData}, designed to retain LSB features and produce visually appealing (but still scientifically accurate) images.
This ERO processing started with the raw data and was independent of the pipelines that have been developed by the VIS, NIR, and MER teams and implemented by the SGS.
This was convenient for processing the 17 ERO pointings, amounting to roughly one day of observations. However, for regular survey operations, it is essential -- for efficiency, consistency, and reproducibility -- to build upon the work of the instrument teams and the SGS.
Our LSB data processing therefore starts with the level 2 (LE2) SGS-processed data products produced by the VIS and NIR processing pipelines.\footnote{These Q1 data products are publicly available from the Euclid Science Archive (\citealt{2017IAUS..325..385D}, \url{http://easidr.esac.esa.int}).}
We take the same individual LE2 flat-fielded and flux-calibrated frames that are used to create the MER tiles, but perform our own LSB-focussed post-processing steps to produce images of our targets of interest, namely galaxy clusters.
We make our code, known as \textsc{nicl}, available\footnote{\url{https://github.com/IntraClusterLight/nicl_q1}} so others may benefit from our LSB processing for specific targets, but we do not intend to provide a complete tiling of the survey. We note that the subsequent DR1 version of the SGS processing features a number of improvements, particularly regarding the NIR background subtraction. However, the need for dedicated post-processing remains for studies of sources larger than one arcminute.

Unfortunately, our post-processing is not as straightforward as simply combining the individual calibrated frames without performing any background subtraction.
As mentioned previously, the MER background subtraction conveniently solves a variety of issues which produce background variations within and between detector chips, and across the FoV.
To produce a high-quality product that preserves flux from the diffuse sources we wish to study, we must take alternative approaches to dealing with these problems.
\refSec{processing} summarises our post-processing steps; \refsec{performance} quantifies their combined impact on the noise properties and faint-source recovery of our images; \refsec{measurement} describes how we mask sources, estimate the background, and measure ICL surface-brightness profiles; and \refsec{simulations} uses mock clusters from numerical simulations to establish the accuracy of our measurements.

We have developed our approach using the Euclid Quick Data Release 1 (Q1; \citealt{Q1cite,Q1-TP001}), and this work describes the data used for the Euclid Collaboration's Q1 studies of ICL
(\citealt[][hereafter K26]{Q1-SP007}; \citealt[][hereafter GM26]{Q1-SP008}; Euclid Collaboration: Gully et al. in prep., hereafter G26). Q1 comprises \qty{63}{deg^2}, located in the Euclid Deep Fields (EDFs), but with nominal EWS depth.
All tests and comparisons in this paper are based on Q1 products.
Although some aspects of the VIS, NIR, and MER processing will change for the first large-scale data release (DR1), the observing strategy for the EWS is identical and we expect that many of the issues we have addressed will remain. We will therefore base the processing for our DR1 ICL studies on that described here.
We summarise the current status of our processing and measurement procedures, and potential further improvements, in \refsec{summary}.

\section{\label{sc:processing}Processing steps}

\begin{figure*}[tbp]
\centering
\includegraphics[width=\hsize]{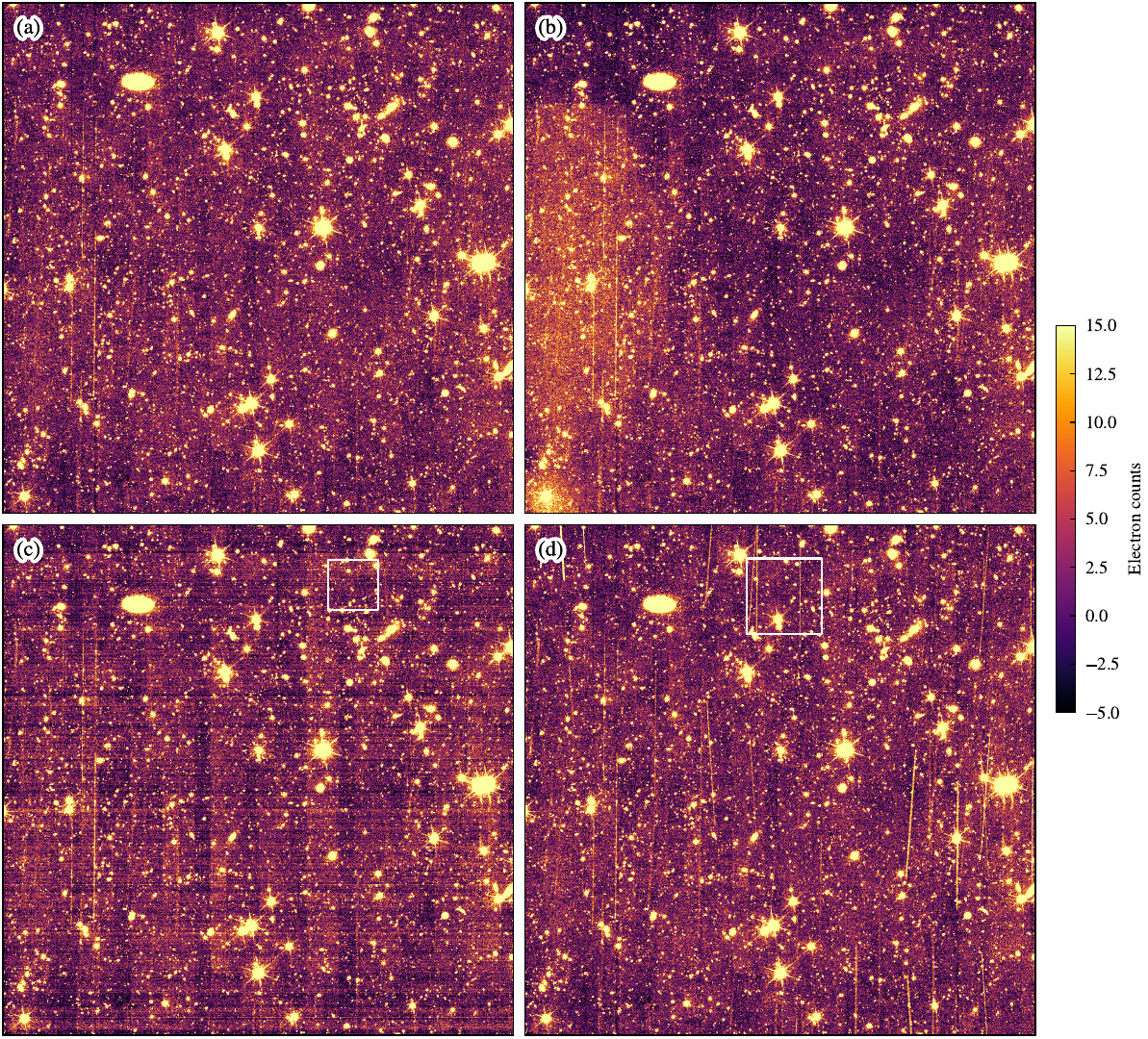}
\caption{\label{fig:single_chip_processing}Single $\num{2040} \times \num{2040}$ pixel detector from a NIR LE2 calibrated frame (a) with our full post-processing, (b) without the atemporal background correction (\refapp{autodark}, the correction itself is shown in the DET11 panel of \reffig{autodark_NIR}), (c) without the tartan correction (\refapp{tartan}, the correction itself is shown in \reffig{tartan_example}), and (d) without the persistence correction (\refapp{persistence}, the correction for the indicated region is illustrated in \reffig{persistence_zoom}).}
\end{figure*}

Our processing starts with the standard LE2 data products. Each calibrated frame (individual exposure) is stored as a multi-extension FITS image, in which each extension contains either a science image, a noise map, or pixel flags of an independently calibrated detector unit. For VIS, this unit is a quadrant of a CCD chip; for NIR, this refers to a whole detector chip. For simplicity, we refer to these as detector images for both throughout this paper. Each EWS observation follows a reference observing sequence (ROS) comprising four dithered nominal-depth exposures in each band (\IE, \YE, \JE, and \HE). The VIS ROS also contains two much shorter \IE exposures that are not used in this work, since they do not contain useful signal for our faint objects of interest and would further complicate our processing algorithms.

The processing steps are described in detail in \refApp{processing}. Here we summarise each step. \refFig{single_chip_processing} illustrates the effect of the first three corrections on a single NIR detector image.
Applying the full post-processing to all observations in Q1 takes only a few days, utilising 16 cores of a modern server. Since the task can be parallelised at the observation level, it is easy to accelerate with high performance computing.

\paragraph{Atemporal background (\refapp{autodark})}
Each NIR and VIS detector exhibits persistent background structure -- including offsets between detectors and within-chip gradients -- that is not fully removed by the LE2 calibration. We construct a per-observation atemporal background model by taking the median of a group of neighbouring observations, then subtract it from each frame. The background structure for NIR and VIS detectors is illustrated in Figs.~\ref{fig:autodark_NIR} and \ref{fig:autodark_VIS}.

\paragraph{Tartan pattern (\refapp{tartan}; NIR only)}
The NIR frames additionally show row- and column-level background variations that change between exposures and are therefore not captured by the atemporal background. We model these by computing the median background separately along each row and column, after masking sources and filtering out large-scale structure, then subtract the resulting 2D model. An example is shown in Figs.~\ref{fig:tartan_projections} and \ref{fig:tartan_example}.

\paragraph{Persistence (\refapp{persistence}; NIR only)}
Bright sources in preceding NISP exposures leave persistence features in subsequent frames. We estimate the persistence signal from temporally-neighbouring exposures and scale it to the target image using an empirically calibrated power-law decay model, then subtract it. Pixels with a large correction uncertainty are masked. An illustration of the correction is shown in \reffig{persistence_zoom}.

\paragraph{Image resampling (\refapp{resampling})}
We use \swarp{} \citep{Bertin+2002} to resample all processed detector images for a given target onto a common \qty{0.3}{arcsec.pixel^{-1}} grid, producing individual resampled images ready for background matching and combining.

\paragraph{Background matching (\refapp{matching})}
Residual background discontinuities at detector boundaries are removed by solving for a constant additive correction per detector image, which minimises differences in the background levels across all overlapping dither pairs.

\paragraph{Image combining (\refapp{combining})}
The resampled, corrected images are combined using \swarp{} in weighted-mean mode with sigma-clipping, to reject residual artefacts.

\begin{figure*}[tbp]
\centering
\includegraphics[width=\hsize]{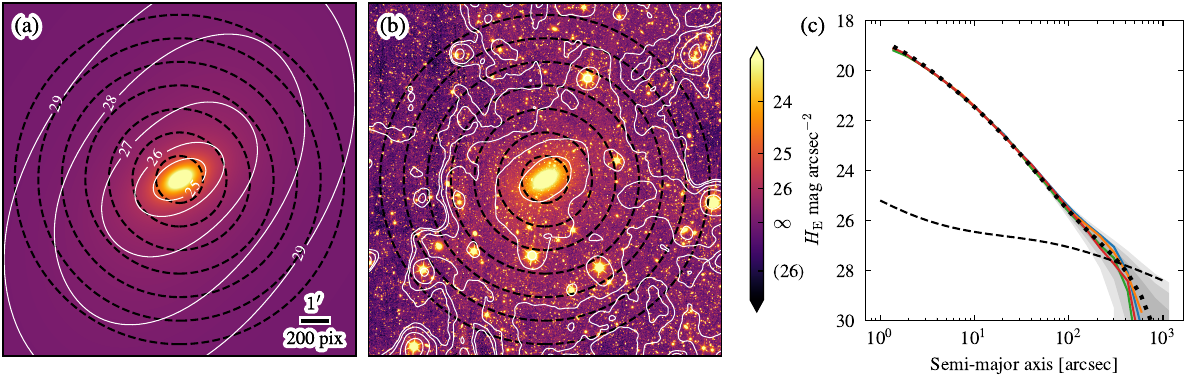}
\caption{\label{fig:mock_test}
Demonstration that our processing preserves faint flux around extended objects. We create a simple mock image of a cluster at $z \sim 0.1$, displayed in the \emph{left} panel. White contours indicate the surface-brightness levels. The isophotal twist is due to an intentional misalignment of the BCG and ICL position angles. The black dashed circles have radii in multiples of $\qty{200}{pixels} = \qty{1}{arcmin}$. The colour scale matches that of the upper row in \reffig{nicl_vs_MER}. The \emph{central} panel contains an image of the mock cluster injected into the \changed{\HE} LE2 calibrated frames and fully processed. Contour levels are the same, but a \qty{1}{arcmin} diameter median filter has been applied to the data before contouring. In the \emph{right} panel, the dotted black line shows the surface-brightness profile measured in the original, noiseless, mock image. Solid coloured lines show the profile for each of the four mock clusters, measured following the procedure described in \refsec{measurement}. The dashed black line indicates the $3\sigma$ background level uncertainty estimated by the procedure described in \refsec{background_noise}. The mock ICL profile is accurately recovered all the way to this limit. \changed{The grey bands around the true profile show the equivalent 1, 2, and 3$\sigma$ confidence intervals corresponding to the background noise.}
}
\end{figure*}

\begin{figure*}[tbp]
\centering
\includegraphics[width=\hsize]{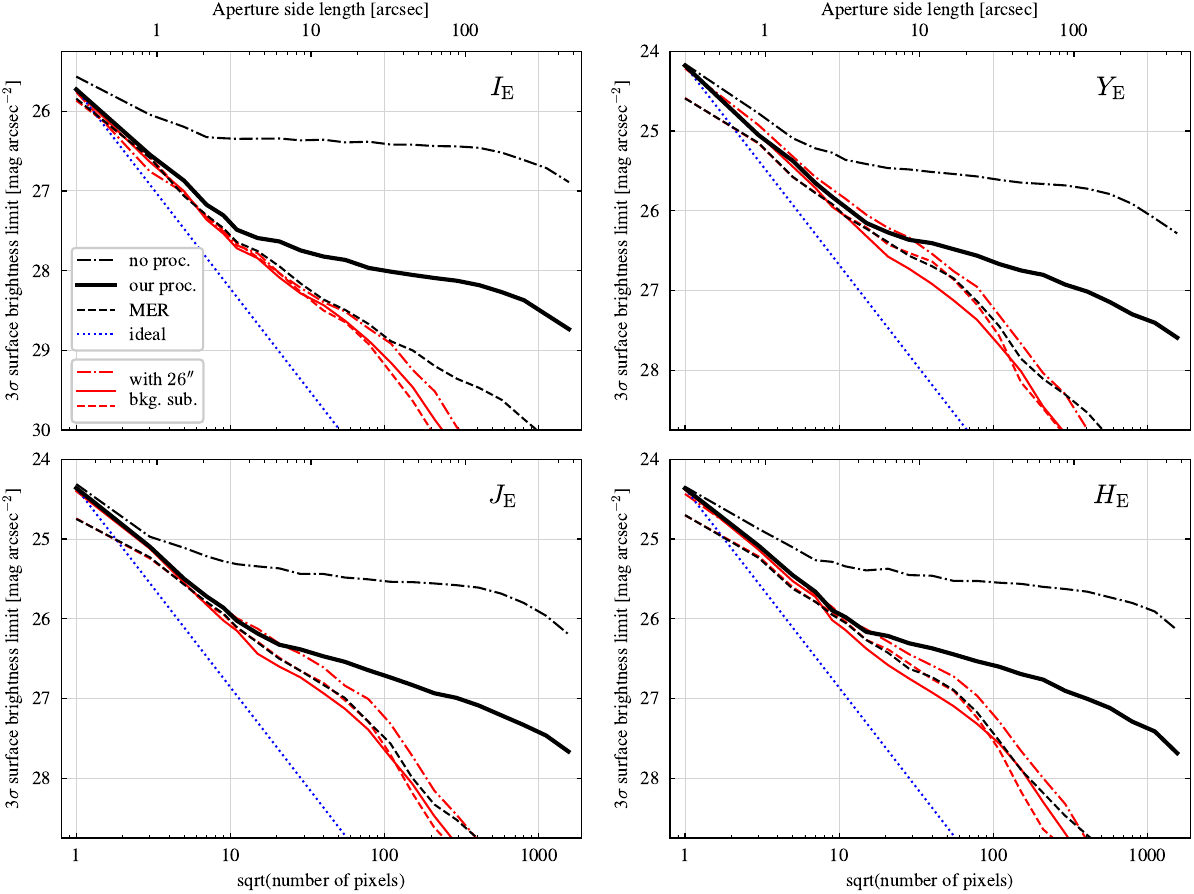}
\caption{\label{fig:noise_in_boxes}
Comparison of the noise properties of our post-processed images (thick, black, solid line) versus not applying any processing (black, dot-dash line) and the official MER processing (with the MER-determined background restored; black, dashed line). We present the $3\sigma$ uncertainty in \unit{mag.arcsec^{-2}}. This is the faintest surface brightness which we could confidently detect in an aperture of a given size. The noise level is estimated from the standard deviation of the median flux measured in many square apertures randomly placed on an image with all sources thoroughly masked (as described in \refsec{flux_retained}). It therefore reflects variations in the background level on the scale of the aperture, since these dominate over the averaged noise on the individual pixels (see \refsec{sb_limit}). The blue, dotted line illustrates the trend expected in the ideal case, where the single-pixel measurement (in our processed images) is extrapolated assuming the noise decreases with the reciprocal square root of the number of pixels.
For all three varieties of data, we have subtracted a large-scale background, estimated in boxes of $\ang{;9} \times \ang{;9}$ followed by a $3 \times 3$ median filter over the boxes. Lines are only shown where the apertures are smaller than this background box size. Our post-processing dramatically improves the limiting surface brightness while retaining extended, faint sources (see \reffig{mock_test}).
In large apertures, while MER appears to have low noise levels, the data is not usable in practice, since faint sources on these scales have been subtracted away (see \reffig{nicl_vs_MER}).
The standard MER products include an additional background subtraction estimated in $\ang{;;26} \times \ang{;;26}$ boxes, followed by a $3 \times 3$ median filter over the boxes, which we have restored here. To illustrate the effect of this, and facilitate a fairer comparison with our processing, in red we show the same set of lines as before, but after subtracting the background estimated on a \ang{;;26} scale. For VIS \IE, our post-processing matches the statistics of MER, while for NIR it appears to provide a slight improvement for extended sources.
}
\end{figure*}

\section{\label{sc:performance}Impact of post-processing on ICL and limiting surface brightness}

\subsection{\label{sc:flux_retained}\changed{ICL flux recovery}}

Each stage of our post-processing additively modifies pixel levels, or masks regions, in order to reduce artefacts. Although these have been designed to avoid removing flux from real sources in the target image, it is important to check this. To do so, we create a simple mock source, shown in \reffig{mock_test}\,(a), mimicking the brightest central galaxy (BCG) and ICL of a cluster at $z \sim 0.1$ (based on the models used by \citealt{EP-Bellhouse}). Four copies of this source, centred on the corners of a $24\arcmin \times 24\arcmin$ square, are injected (following the procedure of \citealt{EP-Urbano}) into the LE2 calibrated frames for a set of 11 observations covering a $40\arcmin \times 40\arcmin$ region of the EDF-S. It is important to locate the mock clusters sufficiently far apart that they do not influence the measurements of each other. We then run our post-processing steps to produce a $30\arcmin \times 30\arcmin$ image of each cluster, one of which is shown in \reffig{mock_test}\,(b). Finally, we run our ICL measurement process (described below in \refsec{measurement}) to extract the profile of each mock cluster. \changed{All bands show the same behaviour, so we show results only for \HE, since this is our preferred band for studying ICL \citep{Q1-SP007}. The example shown includes subtraction of the local background, which is estimated in boxes of \qty{9}{arcmin} (corresponding to \qty{1}{Mpc} at $z = 0.1$), followed by a $3 \times 3$ median filter over the boxes. \changed{This matches the background scale we use in our Q1 ICL studies. Using a much larger background scale only slightly increases the scatter at large radii, as quantified in \refsec{sb_limit} and \ref{sc:background_noise}.}}

\refFig{mock_test}\,(c) demonstrates that the measured profiles accurately follow that of the injected model, within the scatter arising from different background realisations. This continues all the way out to approximately a radius of $\qty{7}{arcmin}$ and surface brightness of $\qty{28}{mag.arcsec^{-2}}$, below the level at which we expect residual background variations (determined below in \refsec{background_noise}). This confirms that our processing has no negative impact on the measurement of the ICL, or other extended, diffuse sources on this scale.

\subsection{\label{sc:sb_limit}Surface brightness limit}

We next investigate the photometric noise levels of our post-processed images. For studies of extended sources, it is common to quote a limiting surface brightness, that is, the faintest source level that one could securely detect (at a given confidence level) above the noise. For faint sources, this limit depends on uncertainties in the flux measurements of individual pixels, correlations between neighbouring pixels (due to detector issues and resampling), and residual spatial variations in the background level of the images. It is therefore important to specify the spatial scale to which the surface-brightness limit corresponds. Furthermore, it is critical not to extrapolate measurements of the noise at the scale of individual pixels to larger sources, but instead to measure the noise on the spatial scale of scientific interest.

There are various ways of quantifying the limiting magnitude. Here we take the approach of measuring the flux in many apertures distributed across a representative region. The scatter on these measurements gives a direct statistical estimate of the uncertainty for a source with a surface brightness comparable to the background level. In \refsec{background_noise} we quantify this uncertainty in annular apertures of exactly the size and shape used to perform our ICL measurements. However, to allow easier comparison with other surveys and processing approaches, in this section we determine the surface-brightness limit in square apertures.

We first create a large image of a $\ang{2.5} \times \ang{2.5}$ region in the EDF-S. To eliminate very large-scale background variations, for example due to zodiacal light or scattered light from sources beyond the \Euclid FoV, we apply a background subtraction. We estimate the background using boxes of side \qty{9}{arcmin}, followed by a $3 \times 3$ median filter over the boxes, sufficient to avoid subtracting the ICL of the largest clusters we intend to study (see \refapp{background_subtraction} for details).
We then mask all sources (as described in \refapp{masking}). We randomly select \num{1000} points in this image, excluding a margin around the edge equal to half of the side length of our largest apertures. At each point, we measure the median flux of the unmasked pixels within 23 square apertures with side lengths increasing geometrically from \qty{1}{pixel} to \qty{3215}{pixels}. For apertures with at least 5 unmasked pixels, we apply iterative $3\sigma$-clipping around the median to reduce the influence of faint, compact sources that escape our mask. After accounting for the pixel scale, we thus have a set of measurements of the average \unit{flux.arcsec^{-2}} for each aperture size. To robustly estimate the scatter in each set of measurements, we calculate their median absolute deviation, and rescale to correspond to a standard deviation. This gives our $1\sigma$ uncertainty on measurements of the \unit{flux.arcsec^{-2}} in each aperture size. We multiply these by 3, then convert to magnitudes, to express this uncertainty in terms of a $3\sigma$ limiting surface brightness in \unit{mag.arcsec^{-2}}.

We estimate the surface-brightness limit as a function of aperture size, in both our fully-processed image and an image created by combining the LE2 calibrated frames without any further processing. We also perform the analysis on a MER image covering the same region. For this, before mosaicking the tiles, we restored the additional background that was determined and subtracted by the MER pipeline. However, the background estimated by the NIR and VIS pipelines has still been subtracted prior to creating the MER tiles. The results are shown by the black lines in \reffig{noise_in_boxes}.
Simply combining the calibrated frames without any further processing results in background variations that limit detections, at any scale, to brighter than ${\sim}\qty{26.3}{mag.arcsec^{-2}}$ for VIS and ${\sim}\qty{25.5}{mag.arcsec^{-2}}$ for NIR. Our post-processing dramatically improves on this, by around \qty{1.5}{mag.arcsec^{-2}} for VIS and \qty{1}{mag.arcsec^{-2}} for NIR, without resorting to a small-scale background subtraction.

For individual pixels in NIR images, the apparent noise measured in MER is significantly lower than in our processed images. This is because MER uses a bilinear interpolation kernel when resampling (see \refapp{resampling}), which strongly correlates pixels with their immediate neighbours. On average one would expect a reduction in the measured standard deviation by a factor of $2/3$, corresponding to \qty{0.44}{mag.arcsec^{-2}}, exactly as seen in the NIR panels. We do not see the same offset for VIS, as our \qty{0.3}{arcsec.pixel^{-1}} images undersample the original pixel scale.

The `ideal' lines in \reffig{noise_in_boxes} show an extrapolation of our single-pixel noise level to apertures containing $N$ pixels, naively assuming that it scales as $1/\!\sqrt{N}$. This is what would be expected if the background pixels were entirely independent of one another. Such an extrapolation would suggest a $3\sigma$ surface-brightness limit for a $10\arcsec \times 10\arcsec$ box of around \qty{30.0}{mag.arcsec^{-2}} in VIS and \qty{28.5}{mag.arcsec^{-2}} in the NIR bands. This is not meaningful, since it is over \qty{2}{mag.arcsec^{-2}} fainter than the detection limit that we actually measure on this spatial scale. \citet{EP-Urbano} show that, for realistic dwarf galaxy profiles with half-light radii ${\sim}10$\,arcsec, the detection limit in VIS MER images is actually $26$--$\qty{27}{mag.arcsec^{-2}}$. For our images, the measured $3\sigma$ surface-brightness limit in a $10\arcsec \times 10\arcsec$ box is \qty{27.8}{mag.arcsec^{-2}} in VIS and \qty{26.4}{mag.arcsec^{-2}} in the NIR bands.

For apertures $\la 10$ pixels, the noise in our processed images is within \qty{0.1}{mag} of that of MER (compare the solid and dashed black lines). The surface-brightness limit does decline with a slope that is close to that of the `ideal' line. However, in larger apertures, the surface-brightness limit in our post-processed data declines more gradually, reflecting that the limit on these scales is primarily determined by residual spatial variations in the background level. While the noise levels in MER appear to decline more rapidly, in practice the data is not usable for photometry in apertures of this size. Source flux on these scales is removed when the MER processing subtracts the background estimated by the VIS/NIR pipelines (see \reffig{nicl_vs_MER}).

The standard Q1 MER products include an additional background subtraction, on a scale of \qty{26}{arcsec} (section 3.3 of \citealt{Q1-TP004}). If we also perform a background subtraction on this scale, the surface-brightness limits are shown by the red lines in \reffig{noise_in_boxes}. The behaviour for MER does not change much, illustrating that the background estimated by the SGS Q1 VIS and NIR pipelines already extends to these small scales. In contrast, the behaviour for combined calibrated frames, with and without our post-processing, now closely matches that of MER. This demonstrates that the apparent low noise in MER is indeed a result of the small-scale background subtraction. When we apply the same approach, our post-processed images have the same noise properties as MER, or, in the case of NIR, perform slightly better. This is likely due to our removal of persistence and tartan artefacts, which cannot be dealt with by a background subtraction.

We can compare our measured surface-brightness limit to those determined for the LSB-specific processing of the ERO observations (section 9.3 of \citetalias{EROData}), although it requires some care.\footnote{\citetalias{EROData} follow the method of \citet{Scaramella-EP1} to estimate a theoretical $1\sigma$, $10\arcsec \times 10\arcsec$ aperture surface-brightness limit, defined in asinh magnitudes, of \qty{29.9}{mag.arcsec^{-2}} in \IE and \qty{28.4}{mag.arcsec^{-2}} in \HE. These values are averages across fields with different sets of observations, but correspond roughly to EWS depth. By definition, a $1\sigma$ flux in standard magnitudes has a value $2.5 \log_{10}(\mathrm{e}) \sinh^{-1}(0.5) = \qty{0.52}{mag.arcsec^{-2}}$ higher than an asinh magnitude. To convert to a $3\sigma$ limit (as given in \reffig{noise_in_boxes}) one must subtract $\log_{10}(3) = \qty{0.48}{mag.arcsec^{-2}}$ from a $1\sigma$ limit. Finally, as explained earlier, bilinear interpolation underestimates the single-pixel noise by \qty{0.44}{mag.arcsec^{-2}}. So, the \citetalias{EROData} values converted to a $3\sigma$ $10\arcsec \times 10\arcsec$ aperture surface-brightness limit defined in standard magnitudes are \qty{29.5}{mag.arcsec^{-2}} in \IE and \qty{28.0}{mag.arcsec^{-2}} in \HE. These limits only account for photon noise due to the zodiacal light background, and are in good agreement with the ideal behaviour shown by the blue-dotted lines in \reffig{noise_in_boxes}.}
\changed{Extrapolating their single-pixel noise measurements, \citetalias{EROData} find good agreement with the pre-launch expectations of \citet{Scaramella-EP1}. However, from direct measurements of physical LSB features in the ERO images, they judge that these theoretical limits are optimistic by $0.4$ mag. Applying this adjustment gives $3\sigma$ surface-brightness limits of \qty{28.3}{mag.arcsec^{-2}} in \IE and \qty{26.8}{mag.arcsec^{-2}} in \HE. These values closely match the solid red lines in \reffig{noise_in_boxes}, for our post-processed images after subtraction of a background estimated on small (\ang{;;26}) scales. The statistics in our images are therefore in excellent agreement with the results of \citetalias{EROData}. However, as they note, their limits do not include pointing-dependent background variations such as cirrus, scattered light from bright stars, and the outskirts of large galaxies. Without the small-scale background subtraction, our measurements (black lines in \reffig{noise_in_boxes}) include all of these additional contaminants, and hence give the limiting surface brightness of a $10\arcsec \times 10\arcsec$ source that could be detected, with $3\sigma$ significance, if located at a random position in the Q1 EDF-S: \qty{27.8}{mag.arcsec^{-2}} in \IE and \qty{26.4}{mag.arcsec^{-2}} in the NIR bands.}

We have thus quantified the surface-brightness limit of our data and demonstrated that we can obtain low photometric noise levels without compromising the faint signal from extended sources.

\section{\label{sc:measurement}Measuring ICL}

Full details of the background subtraction, masking, and photometry procedures are given in \refApp{measurement}. Below we summarise the approach. \refFig{sim_test}\,(c) provides an example of our masking and fitted elliptical isophotes, while \reffig{mock_test}\,(c) shows a measured surface-brightness profile. Figure 2 in \citetalias{Q1-SP007} presents an example for a real cluster.

\subsection{Background subtraction}
Our processing so far has avoided performing a traditional background subtraction. Although we have attempted to remove several sources of background variation, primarily those associated with the detectors and camera optics, other sources will remain. These include the extended PSF of very bright stars, scattered light from outside the FoV, zodiacal light, Galactic cirrus, and even clustering in the faint unresolved galaxy population \citep{2023MNRAS.525.1443L}. When measuring the photometry of specific objects, even the outskirts of other large clusters and galaxies may be considered background. We therefore need to perform a background subtraction, while ensuring that it does not negatively impact our measurements of the ICL.

We use \texttt{photutils.Background2D} \citep{photutils}, following the \texttt{SExtractor} approach \citep{1996A&AS..117..393B}, with a background box size corresponding to \qty{1}{Mpc} at the cluster redshift (e.g., \qty{9}{arcmin} at $z=0.1$), followed by a $3 \times 3$ median filter. This scale is large enough to preserve the ICL signal of all clusters used in our initial Q1 ICL studies (\citetalias{Q1-SP007}; \citetalias{Q1-SP008}; G26). See \refapp{background_subtraction} for details.

\subsection{Masking}
When performing photometry of the ICL, we also need to eliminate contamination by other sources, namely stars and galaxies. While more sophisticated solutions exist (e.g., \citealt{2018ApJ...857...79J, 2021A&A...649A..38E}), here we take the simplest approach, of masking. We therefore construct a mask containing the pixels that are likely to contain significant flux from contaminating sources. We also use such a mask to improve our estimate of the background.

As detailed in \refapp{masking}, we construct three masks: a bad-pixel mask, a regular-objects mask (sources outside the ICL region, grown iteratively to 50\% larger than their detected size), and a central-objects mask (compact sources within the ICL region, grown to 25\% larger). The union of these three forms the measurement mask. The background mask additionally includes an ICL mask, which delineates the extent of the diffuse cluster light via an adaptive smoothing algorithm. For NIR, a single mask derived from a combined \YJH image is used consistently across all NIR bands.

\subsection{Photometry}
For our initial Q1 ICL studies (\citetalias{Q1-SP007}; \citetalias{Q1-SP008}; G26), we take a straightforward approach to photometry using elliptical isophotes. We then quantify the most important source of uncertainty, namely variations in the background level, as described in \refsec{background_noise}.

Elliptical isophotes are fit using \texttt{AutoProf} \citep{2021MNRAS.508.1870S}, with fixed centres and semi-major axes, optimising ellipticity and position angle (PA) at each annulus. Flux in each annulus is measured as the masked, sigma-clipped median using \texttt{photutils.aperture}. For multi-band colour profiles, isophotes are defined on a combined \YJH image and applied to all bands, including \IE. Details are given in \refapp{photometry}.

\begin{figure}[tbp]
\centering
\includegraphics[width=0.975\hsize]{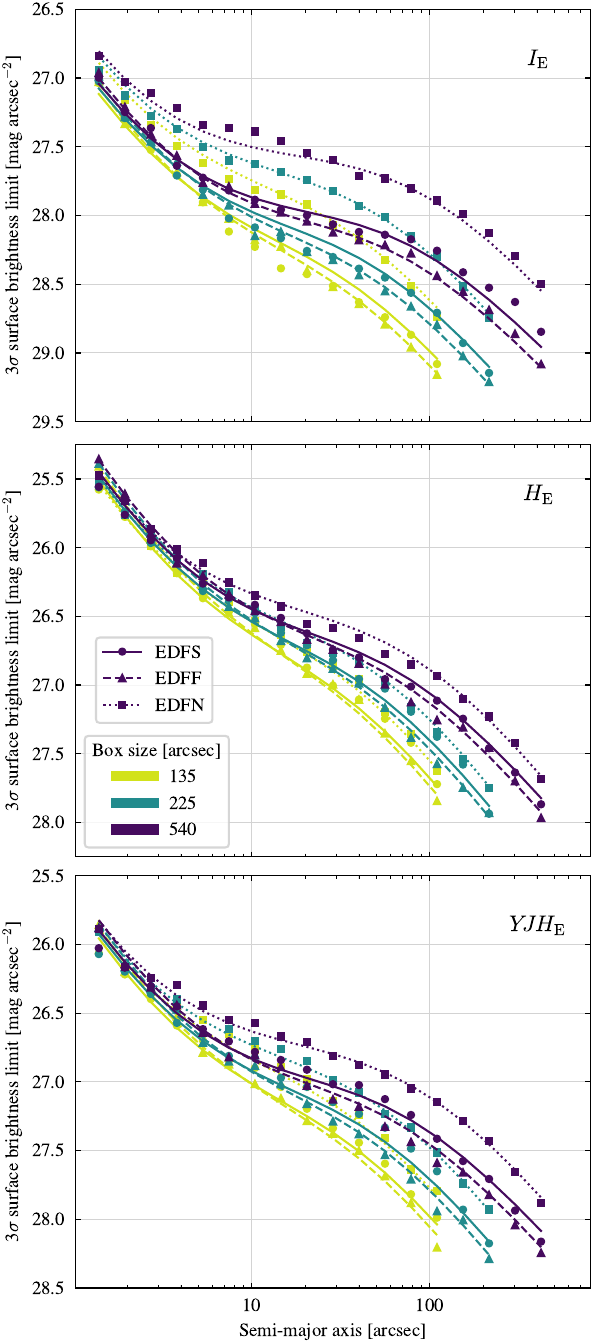}
\caption{\label{fig:sky_uncertainty}
$3\sigma$ uncertainty in the background level for elliptical annular apertures of varying central semi-major axis, $r_\textrm{c}$, width $r_\textrm{c}/3$ and ellipticity of $0.5$. Points show the measurements, while the lines correspond to an analytic model. The uncertainty is estimated from the standard deviation of the median flux measured in many apertures randomly placed on a $\ang{2.5} \times \ang{2.5}$ image with all sources thoroughly masked (as described in \refapp{masking}). It therefore reflects variations in the background level on the scale of the aperture, since these dominate over the averaged noise on the individual pixels (see \refsec{sb_limit}). In the figure, we express the $3\sigma$ uncertainty in \unit{mag.arcsec^{-2}}. We use this as the surface brightness above which we are able to obtain a reliable measurement of the ICL profile. Only results for \IE, \HE, and \YJH are shown, as the models for \YE and \JE are everywhere within \qty{0.1}{mag.arcsec^{-2}} of those for $\HE$.
}
\end{figure}

\subsection{\label{sc:background_noise}Background level uncertainty}

\refSec{sb_limit} explains how we estimate a surface-brightness limit due to residual background variations in our processed images. In order to correctly quantify the uncertainty on our BCG+ICL surface-brightness profiles, we perform the same analysis using the elliptical annular apertures defined in \refapp{photometry}. The ellipticity of the apertures is fixed to 0.5, which is the typical value we measure in real clusters \citepalias{Q1-SP007}. The PA is varied randomly for each centre position.

We produce a $\ang{2.5} \times \ang{2.5}$ image for each of the three Q1 fields. For each we perform a background subtraction with 14 box sizes covering the range \qty{1.5}{arcmin} to \qty{14}{arcmin}. For each combination of field, background box size, and annulus semi-major axis, we use \num{1000} apertures to estimate the standard deviation of the measured background flux. The resulting uncertainties, expressed in terms of surface brightness, are plotted in \reffig{sky_uncertainty}. The curves display similar behaviour, with systematic differences for VIS versus the NIR bands.

There is a higher prevalence of Galactic cirrus in the EDF-N, located at Galactic latitude $b = \ang{30}$ with $E(B-V) \approx 0.06$, compared to EDF-S and EDF-F, at $b=\ang{-47}$ and \ang{-55} with $E(B-V) \approx 0.02$ and $\approx 0.01$, respectively (see figure 1 of \citealt{Q1-TP001}). Our measured surface-brightness limits in each field vary in a corresponding, systematic manner. On arcminute scales, the limit in the EDF-N is brighter than the other two fields by around \qty{0.4}{mag.arcsec^{-2}} in \IE and \qty{0.2}{mag.arcsec^{-2}} in \HE. The EDF-S limit is also slightly brighter than that for EDF-F, around \qty{0.1}{mag.arcsec^{-2}} in \IE and \qty{0.05}{mag.arcsec^{-2}} in \HE. This indicates that, even in fields far from the Galactic plane and particularly for EDF-N, the surface-brightness limit on large scales is determined by the presence of cirrus.

To smooth these measurements, and interpolate them to any choice of box size, we perform symbolic regression using \texttt{PySR} \citep{2023arXiv230501582C}. This produces an analytic model, with three coefficients describing how the function varies for each band, and another three coefficients used to describe the variations for each field. As shown in \reffig{sky_uncertainty}, the lines capture the range of behaviour of the surface-brightness limit, with a residual RMS of \qty{0.003}{mag.arcsec^{-2}}.

The uncertainty in our BCG+ICL photometry, particularly for the largest radii at which we can detect ICL, is dominated by the uncertainty on the background level quantified above. It is only within the centre of the BCG that sources of noise on individual pixels (e.g., photon counts and detector readout noise) become important. As mentioned in \refapp{photometry}, we quantify the uncertainty on the latter using the standard error of the pixel values within each annulus. This is combined in quadrature with the uncertainty on the background level, to give an empirical estimate of the total uncertainty in each annulus.

\begin{figure*}
\centering
\includegraphics[width=1\hsize]{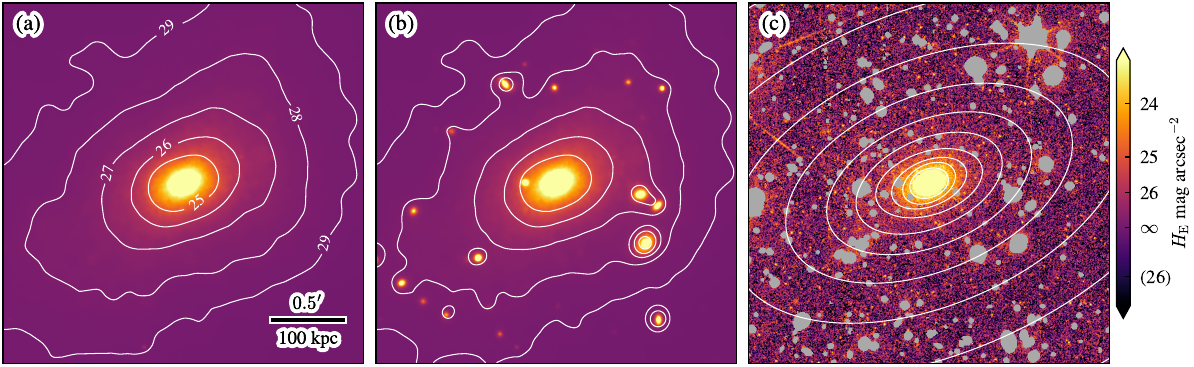}\\
\caption{\label{fig:sim_test}
Example of one of the mock clusters, at $z=0.2$, created from the Hydrangea simulation. We use these to confirm our ability to accurately recover the flux and shape profile of the BCG+ICL. Panel (a) shows the ideal mock image of just the BCG+ICL light, whereas (b) additionally includes satellite galaxies. Surface-brightness contours are shown in these panels; these have been smoothed on a scale of \qty{10}{arcsec}. Panel (c) shows the data from (b) injected into a fully-processed \Euclid image. Grey regions indicate masked sources. The curves show the elliptical isophotes fit using \texttt{AutoProf}, as part of our measurement procedure. The colour scale is the same as for the upper row of Figs.~\ref{fig:nicl_vs_MER} and \ref{fig:mock_test}.}
\end{figure*}

\section{\label{sc:simulations}Testing ICL measurements with realistic clusters}

Above, we have shown that our post-processed cluster images are suitable for measuring faint, extended sources (\refsec{performance}), and explained how we perform our measurements of the surface-brightness profile and shape of the BCG+ICL (\refsec{measurement}). We now explore how accurately our observational measurements represent the true values of these properties. To do so, we construct a set of mock images of simulated clusters, then analyse these using the same methods we apply to real data.

\subsection{Mock images of clusters}

Mock images are generated from the Hydrangea cosmological hydrodynamical simulations \citep{2017MNRAS.470.4186B}, following the approach of \citet{2022MNRAS.513.1459M}. We construct 776 mock cluster configurations at six target redshifts ($z \approx 0.1$--$0.8$), spanning a mass range $M_{200} = 10^{14}$--$10^{15}\,\Msun$. We do not expect the results of these tests to depend significantly on the observed filter, so we focus only on the \HE band. Full details are given in \refApp{simulations}.

For each projected cluster we construct an `ideal' image, containing only the BCG+ICL without noise or background. We use this to determine the `true' projected shape and surface-brightness profile of the BCG+ICL, by analysing it with the measurement steps described in \refapp{photometry}, omitting the background subtraction and masking. The centre is fixed to the known cluster centre, and the background is fixed to zero. Only the PA and ellipticity of the isophotes are allowed to vary. The semi-major axes of the isophotes are defined as for the real data. Since these images contain only the smooth BCG+ICL component, and are free of satellites and observational noise, they provide the reference profiles against which we can compare other measurements.

Observationally `realistic' images of the mock clusters, both with and without satellites, are created by injecting them into fully-processed \Euclid images from a region chosen to be free of real clusters. The resulting images therefore contain noise, artefacts, large-scale sky variations, non-cluster galaxies, and stars, thereby closely reproducing the conditions under which we perform our measurements for real clusters. \changed{This section assumes that our images preserve diffuse light signal, as discussed in \refsec{flux_retained}, down to the limiting surface brightness quantified in \refsec{background_noise}.} The images are analysed with our full measurement pipeline (which includes all the steps described in \refsec{measurement}). \changed{This ensures that observational systematics and choices of our measurement method affect the mocks in the same way as the real data.}

 \begin{figure}[tbp]
\centering
\includegraphics[width=\hsize,clip,trim=1.25cm 0.5cm 1.0cm 0.5cm]{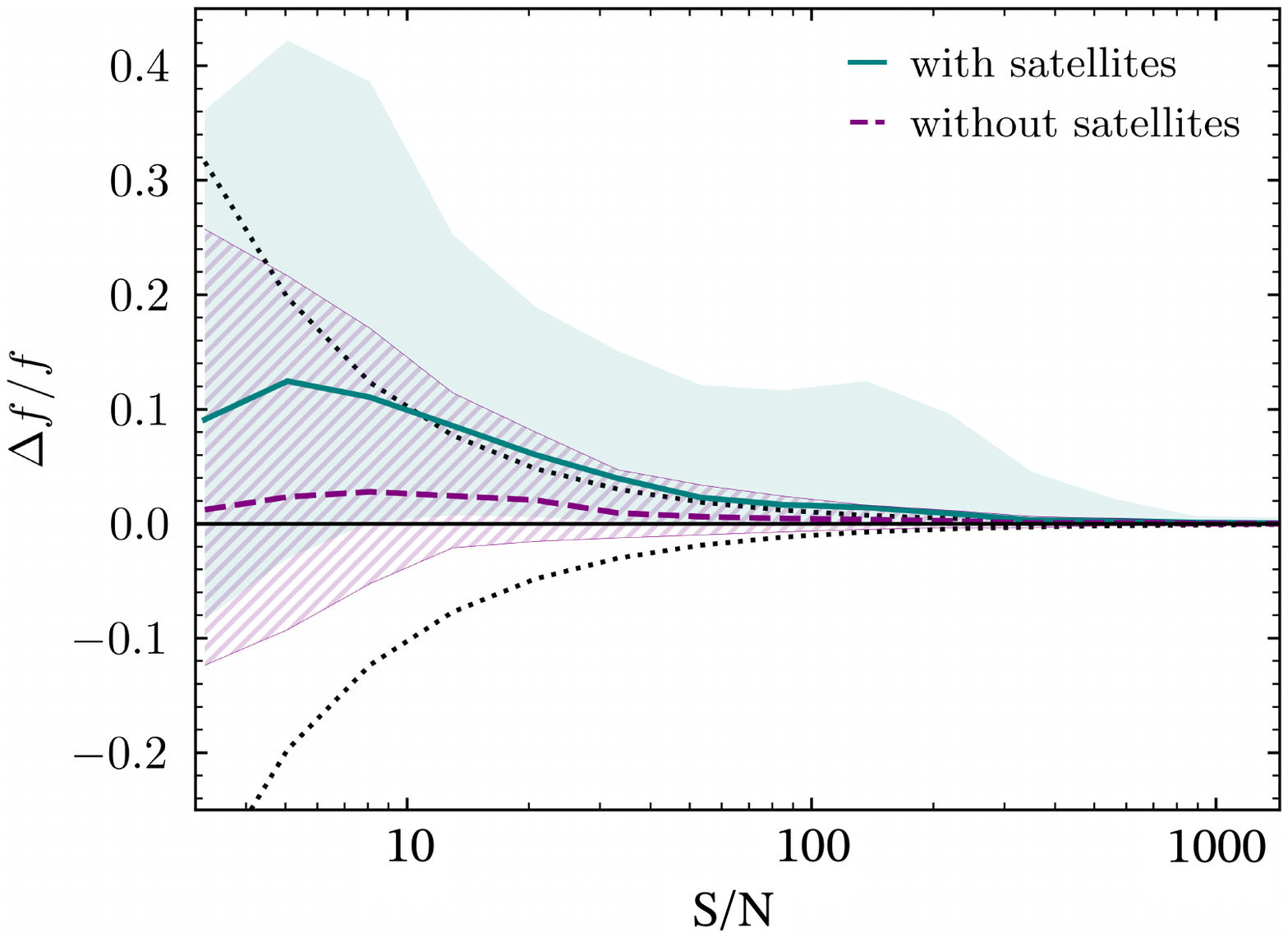}
\caption{\label{fig:flux}
Fractional difference, $\Delta f / f$, between the flux measured in the realistic mock images and that from the ideal BCG+ICL images, plotted as a function of the signal-to-noise ratio (\snr) in the annular aperture with which the measurement is made. For this comparison, the apertures are defined on the ideal image. The purple dashed line shows the median in bins of \snr for realistic mock images in which only the BCG+ICL light has been injected, omitting subhaloes containing satellite galaxies. The region shaded with diagonal lines shows the corresponding 16--84th percentile region. The blue solid line, and smoothly shaded region, shows the same for the case where subhaloes are included.
The dotted black lines trace $\pm (\snr)^{-1}$ around zero.
As discussed in the text, the levels of bias and scatter can be well understood, and indicate that we are able to measure the surface brightness of the ICL with a systematic uncertainty of around $10\%$.
}
\end{figure}

\subsection{Recovering the BCG+ICL flux}
\label{sc:sim_flux}
We start by studying the fractional difference between the flux profile measured in the realistic mock images and that from the ideal BCG+ICL images. For this comparison we use forced photometry, so the comparison is not affected by variations in the shape of the elliptical annuli. The elliptical annuli derived from the ideal BCG+ICL-only images are used to measure the surface-brightness profile in the realistic mock images. Since the brightness and size of the clusters vary, we present the fractional flux difference as a function of the signal-to-noise in the aperture, in \reffig{flux}. The best case would be that we find an average of zero, with a scatter given by $(\snr)^{-1}$.

We first consider the results when satellite galaxies are excluded from the realistic images (purple line). For apertures with $\snr \ga 100$, the scatter around zero is below $2\%$, so there is essentially no difference between the measured and true flux. Towards lower \snr, corresponding to the outskirts of the ICL for any given cluster, there is an increasing scatter about zero. The overall scatter is similar to the expected $(\snr)^{-1}$, indicating that it is primarily due to the background variations quantified in \refsec{background_noise}. However, the asymmetry of the scatter, and the median line, indicate a small bias such that the median flux measured on the realistic images is overestimated by 3\% in the lowest \snr annuli. This is presumably due to faint flux from the outer regions of projected galaxies being included in our measurements. The scatter around the median line is roughly $10\%$ below and $20\%$ above, at our detection limit of $\snr = 3$. The bias is thus small compared with the scatter, so it is negligible for individual clusters, but it would be preferable to account for it when combining the measurements of many individual clusters.

We next investigate the impact of including cluster satellite galaxies in the realistic image. This increases the number of galaxies in the ICL region. Furthermore, due to tidal interactions with other satellites and the cluster itself, these galaxies are likely to be surrounded by extended diffuse light from stars that are being stripped. In a simulation, with full knowledge of the kinematics and three-dimensional positions, we can determine which stars are still bound to the satellite subhaloes. However, there is some flexibility in this definition and it varies with the subhalo finder \citep{2019MNRAS.482.2039C,2025MNRAS.543.1339F}. From \reffig{sim_test} it is apparent that there is some light around satellite galaxies, below the surface-brightness limit of the realistic images, which is not completely masked. However, it is still reasonable to classify this signal as ICL. At this level, therefore, variations in masking amount to choices in the definition of the ICL, analogous to those in the simulations. The optimal definition will depend upon one's particular science goals.

In \reffig{flux}, we additionally show the fractional flux difference for the realistic images containing satellites. There is a strongly asymmetrical scatter, with the flux almost always greater than the ideal BCG+ICL flux, indicating that the definition of the ICL adopted for the simulations excludes more light than our observational choices. The scatter below the median remains close to the expectation given the \snr, whereas above there is an additional contribution to the scatter, which increases from around $5\%$ at high \snr, to roughly $20\%$ in the lowest \snr annuli. This translates to a positive bias on the median of up to $14\%$ at $\snr = 3$. This issue should be kept in mind when comparing observational measurements with simulations, where ICL is defined using 6D information, but does not imply a substantial additional uncertainty when comparing measurements for different clusters.

Even if one prefers the ICL definition adopted in the simulation, the discrepancy rarely leads to an overestimate of more than 50\% in the ICL flux within the outermost annuli. Furthermore, many of these additional stars, if not already part of the ICL, soon will be \citep{2024MNRAS.535.2375M}, and their inclusion should be fairly consistent across our observational cluster sample. We therefore do not expect this level of contamination to significantly affect our interpretation of the measured radial flux and colour profiles. \changed{Nevertheless, as we move to analysing larger samples of ICL measurements, it will be important to better understand the link between ICL measurements in simulations and observations, and its dependence on cluster properties.}

 \begin{figure}[tbp]
\centering
\includegraphics[width=\hsize,clip,trim=1.25cm 0.5cm 1.0cm 0.5cm]{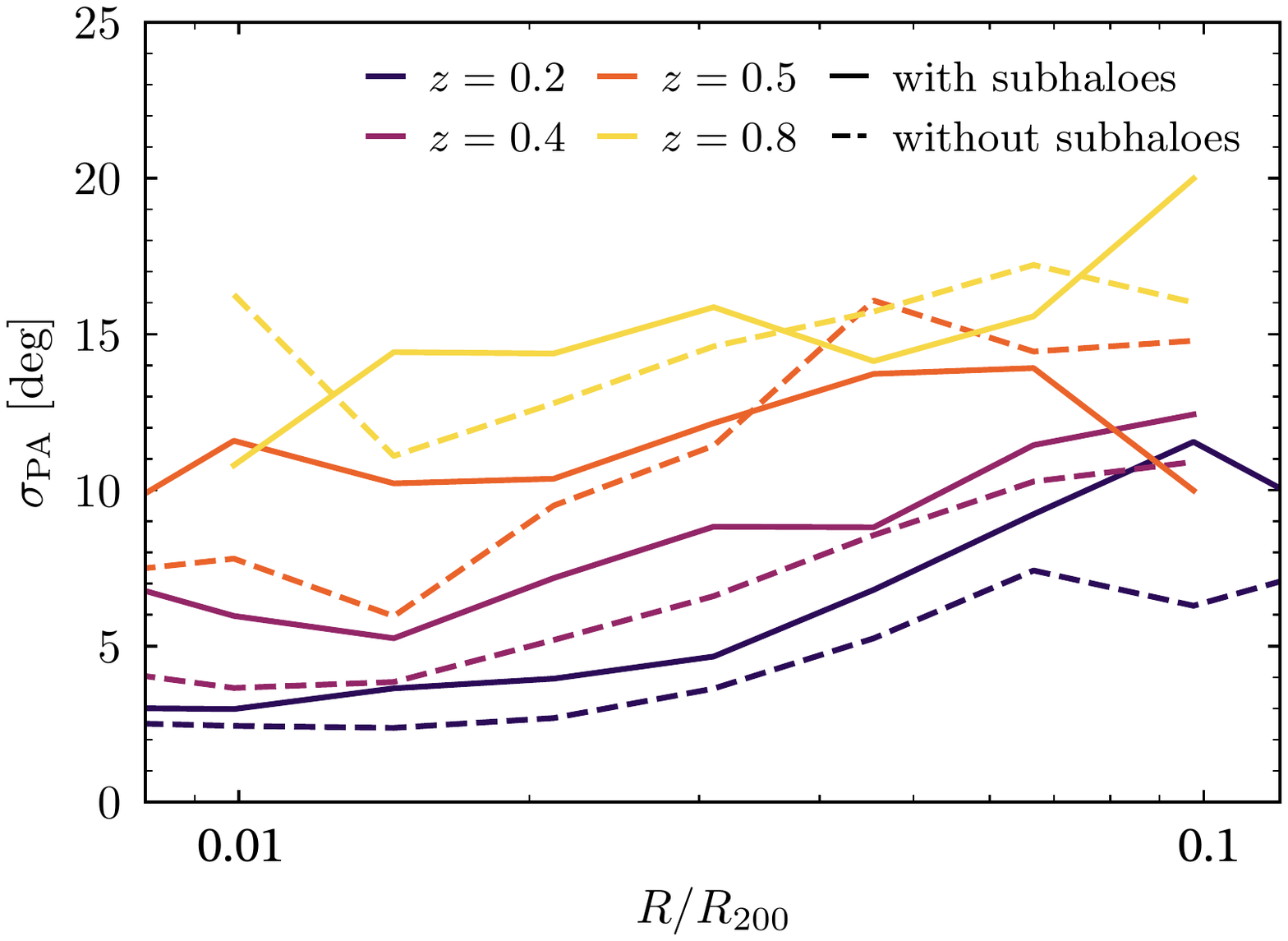}
\caption{\label{fig:PA}
The scatter, robustly estimated via the median absolute difference, on the difference in position angle (PA) measured in the realistic mock images versus the ideal BCG+ICL image. We plot the scatter in bins of the normalised radius ($R/R_{200}$), for a sample of cluster redshifts. Dashed lines correspond to realistic mock images in which only the BCG+ICL light has been injected, i.e. omitting subhaloes containing satellite galaxies. Solid lines show the case when subhaloes are included, which does not particularly change the scatter. The typical uncertainty on our measurement of the ICL PA is around \ang{10}.
}
\end{figure}

\subsection{Recovering the BCG+ICL position angle}
In \reffig{PA} we show the scatter on the difference in the PA measured in the realistic images versus the ideal BCG+ICL image. We use this as an estimate of the uncertainty on our PA measurements. We find that the scatter is dependent on both radius and the redshift of the cluster (which impacts both the angular size of the ICL and the surface brightness). It can be seen that adding the satellites does not particularly change the scatter. We obtain PA uncertainties of \ang{3} for BCGs at $z \sim 0.2$ up to $\ang{15}$ in the ICL at $z \sim 0.8$.

 \begin{figure}[tbp]
\centering
\includegraphics[width=\hsize,clip,trim=1.25cm 0.5cm 1.0cm 0.5cm]{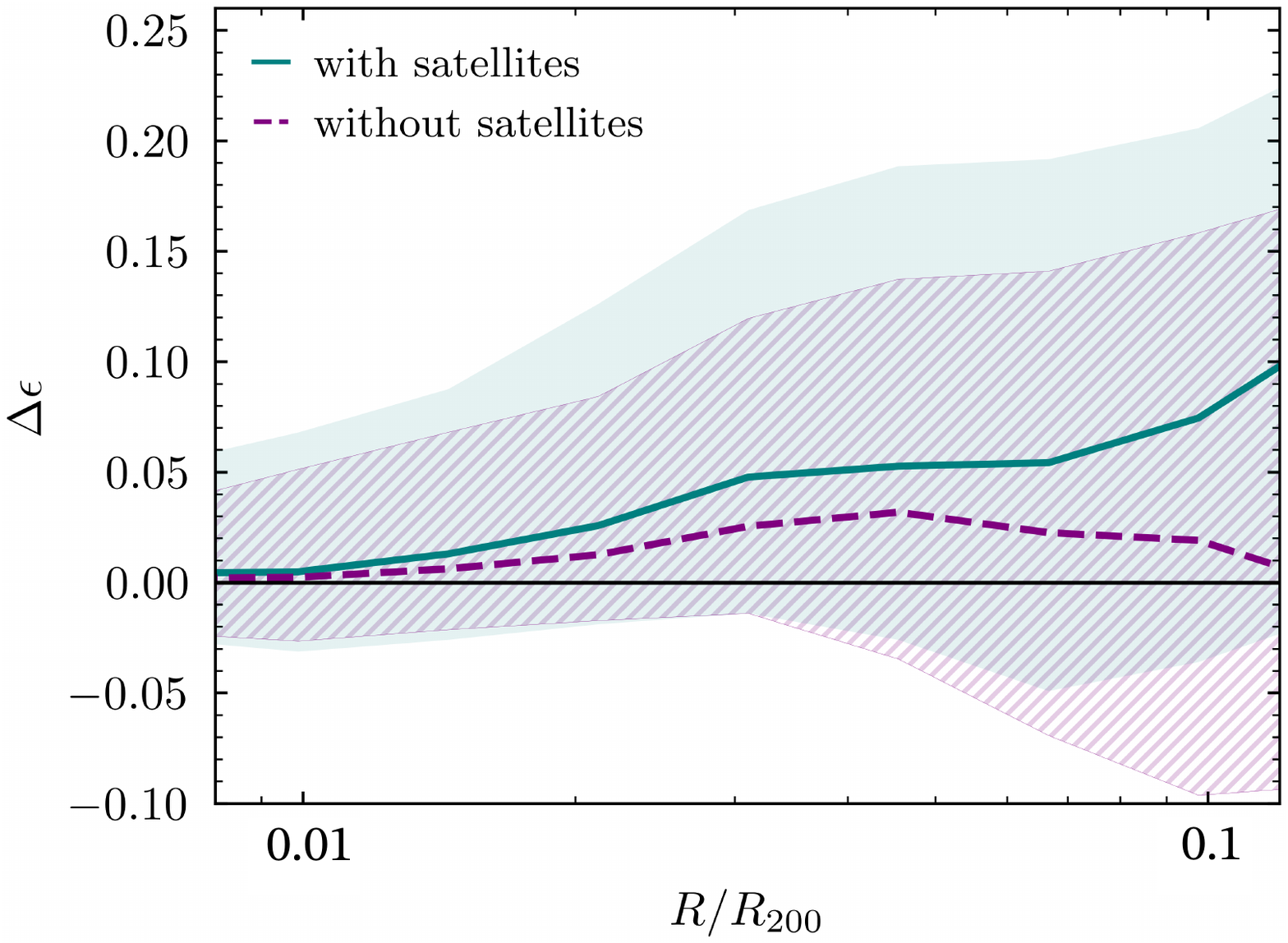}
\caption{\label{fig:ellip}
Difference between the ellipticity $\epsilon$ measured in the realistic mock images and that from the ideal BCG+ICL images. The dashed line shows the median in bins of normalised radius $R/R_{200}$ across all the realistic mock images in which only the BCG+ICL light has been injected, i.e. omitting subhaloes containing satellite galaxies. The region shaded with diagonal lines shows the corresponding 16--84th percentile region. The solid line, and smoothly shaded region, shows the same for the case where subhaloes are included.
As discussed in the text, the levels of bias and scatter can be well understood, and indicate that we are able to measure the ellipticity of the ICL in individual clusters with an uncertainty of around $0.1$.
}
\end{figure}

\subsection{Recovering the BCG+ICL ellipticity}
We now consider the difference in ellipticity between measurements on the realistic images and the ideal BCG+ICL image, as a function of semimajor axis. We use the scatter in this difference as an estimate of the uncertainty on our ellipticity measurements. When doing so we estimate the scatter using only mock clusters in the redshift bin containing the measurement subject. The behaviour is similar for the mock clusters in each redshift bin, so in \reffig{ellip} we display the distribution for all the clusters taken together. \citetalias{Q1-SP007} presents further details of the shape measurements of both real and simulated clusters, and their uncertainties.

Placing the BCG+ICL in the realistic images results in a scatter on the measured ellipticity of $0.04$ in the inner regions, up to $0.12$ at a radius around $0.1\,R_{200}$, where we typically measure the ellipticity of the ICL. There appears to be a small bias in the measured ellipticity, such that the average ellipticity is increased by up to $0.03$. This bias peaks at intermediate radii, where the flux profile is relatively faint, but our measurement annuli are still quite narrow. In this region, limitations in masking the outskirts of projected galaxies may have an impact on the shape of the isophote. The bias is always small compared with the scatter, and declines towards $0.1\,R_{200}$, so it is not significant for measurements of individual clusters. However, it needs to be accounted for when combining measurements of multiple clusters \citepalias{Q1-SP007}.

There is a trade-off between the level of contamination from flux around projected (and satellite) galaxies and the uncertainty on our flux and shape measurements of the BCG+ICL. Masking a larger area around each galaxy would reduce potential contamination, but would leave less area in which to measure the BCG+ICL. The precision of these measurements would therefore decrease. It could also introduce a bias: if satellites are distributed closer to the major axis, the unmasked region will be missing BCG+ICL signal in that particular region. In selecting our masking parameters we have aimed to find a balance that results in the highest accuracy for our BCG+ICL measurements.

When we include satellite galaxies, the scatter remains similar, but there is a significant increase in the apparent bias. At $0.1\,R_{200}$, the ellipticity measured in the realistic mock images with satellites is $0.08$ higher than that from the ideal BCG+ICL images. As discussed in \refsec{sim_flux}, these realistic images include some unmasked flux around each satellite. This will have some influence on the ellipticity as well as the measured flux. If satellite galaxies tend to be distributed along the major axis of the cluster, the additional flux will, on average, raise the measured ellipticity. However, for the same reasons given in \refsec{sim_flux}, at this level one could consider the additional flux to be ICL. The discrepancy is as much due to the definition of ICL in the simulations as it is our masking choices. The optimal definition of ICL will depend upon the science question being addressed.

\section{\label{sc:summary}Summary}

This paper describes the post-processing and measurement procedures used for our studies of ICL with the \Euclid Q1 data release (\citetalias{Q1-SP007}; \citetalias{Q1-SP008}; G26).
The official \Euclid (MER) processing is optimised for weak lensing measurements and removes the signal from diffuse structures. We therefore develop an alternative approach, starting from the calibrated frames for each dithered exposure, produced by the NIR and VIS pipelines. Through a series of additive corrections, we are able to produce images possessing a very flat background level without resorting to performing a local background subtraction. We are therefore able to perform a background subtraction tailored to the spatial scale of individual extended targets.

We quantitatively demonstrate the quality of our images, showing that their noise properties compare favourably with the official MER products and the bespoke processing of the ERO observations. However, in contrast to MER, our approach preserves extended, LSB structure. We show this by accurately recovering the light profile for mock clusters injected into the individual dithers prior to our post-processing. On scales of several arcminutes, we can reliably detect surface-brightness levels below \qty{27.5}{mag.arcsec^{-2}} in \YE, \JE, \HE, and around \qty{29}{mag.arcsec^{-2}} in \IE. At these levels, the most important remaining source of background variations appears to be Galactic cirrus, even in regions far from the Galactic plane.

We describe our approach to measuring the surface-brightness profile and shape of the ICL, including our choices regarding mask creation and background estimation. Applying this to mock images created using numerical simulations, we show that we can obtain accurate measurements of the ICL in individual \changed{$z \approx 0.1$--$0.8$} clusters with uncertainties on the order of 10\% in surface brightness, \ang{10} in position angle, and $0.1$ in ellipticity. We highlight systematic uncertainties due to choices in assigning diffuse flux around satellite galaxies to the ICL, but argue that these can be dealt with by adopting a consistent approach when comparing within observations and between observations and simulations.

The procedures outlined in this paper will provide the basis for our future ICL work with \Euclid, and will also be applicable to the {\it Roman} Space Telescope \citep{2019arXiv190205569A,2024SPIE13092E..0SS}.

Given our promising initial results, we expect to meet the ambitious forecasts of \citet{EP-Bellhouse}, detecting ICL in tens of thousands of clusters in the EWS out to $z \sim 1.5$ \changed{\citep{Q1-Gully}}, with hundreds of these having ICL profiles measured to radii beyond \qty{500}{kpc}. The EDF data, and stacking of clusters from the EWS, will both enable us to push studies of the ICL to even fainter levels. Together, this work will revolutionise our understanding of the ICL, and galaxy cluster haloes in general.

\begin{acknowledgements}

N.A.H., H.Gao, T.K., J.B.G.M., H.Gully, and M.U. were funded by the Leverhulme Trust through RLA-2022-002.
N.A.H. and S.P.B. were funded by the UK Science and Technology Facilities Council (STFC) under grant ST/X000982/1.
Y.M.B. acknowledges support from UK Research and Innovation through a Future Leaders Fellowship (grant agreement MR/X035166/1) and financial support from the Swiss National Science Foundation (SNSF) under project ``Galaxy evolution in the cosmic web'' (200021\_213076).
We are grateful for access to the University of Nottingham's Ada HPC service.
This work made use of Astropy (\url{www.astropy.org}) a community-developed core Python package and an ecosystem of tools and resources for astronomy \citep{astropy:2013, astropy:2018, astropy:2022}.
This research made use of Photutils, an Astropy package for
detection and photometry of astronomical sources \citep{photutils}.
This research made use of AutoProf, a package for galaxy image
photometry \citep{2021MNRAS.508.1870S}.

  \AckQone

  \AckEC

\end{acknowledgements}

\bibliography{aa62256-26}

\begin{appendix}

\section{\label{ap:processing}Post-processing details}

\subsection{\label{ap:autodark}Atemporal background (NIR \& VIS)}

\begin{figure}[tbp]
\centering
\includegraphics[width=\hsize]{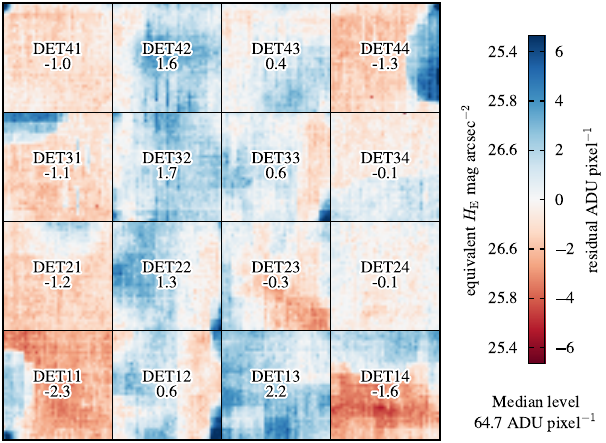}

\caption{\label{fig:autodark_NIR}Example of the NIR atemporal background ($\HE$ band for observation 2683). Detectors are shown as arranged in the camera, but without inter-detector gaps. We note the background level offsets between detectors and significant spatial structure within many of the individual chips. The number appearing below each detector label gives the median residual for that detector.}
\end{figure}

\begin{figure*}[tbp]
\centering
\includegraphics[width=\hsize]{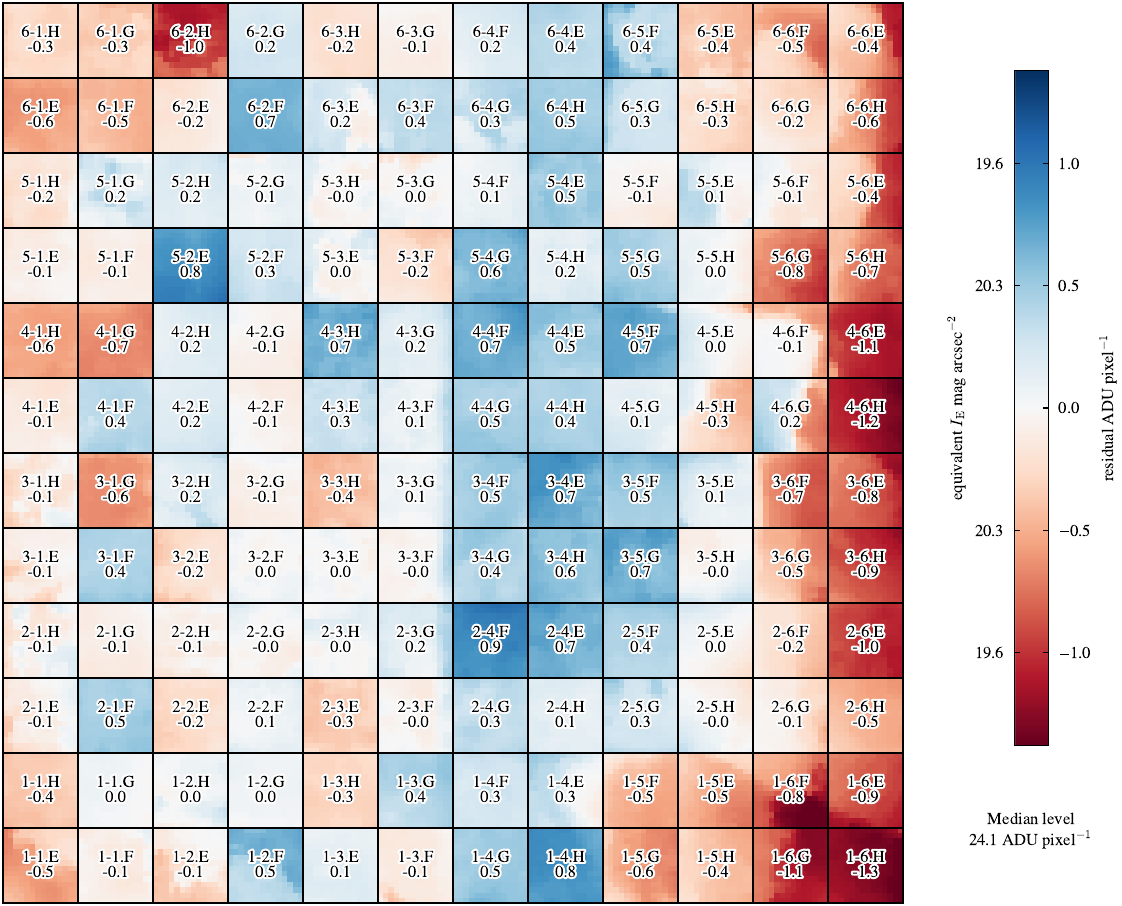}
\caption{\label{fig:autodark_VIS}Example of the VIS atemporal background. Detector quadrants are shown as arranged in the camera, including half-turn rotations, but without inter-detector gaps. The background displays offsets in the average level between neighbouring quadrants, as well as significant structure on larger and smaller spatial scales. The number appearing below each quadrant label gives the median residual for that quadrant.}
\end{figure*}

The LE2 processing of the VIS and NIR data includes steps to remove illumination-independent signal (e.g., bias and dark current) and variations in pixel response \citep{Q1-TP002, Q1-TP003}.
The latter uses LED flats, followed by a large-scale illumination correction based on variations in the measured magnitudes of stars across a 76-point dither pattern.
The resulting correction is determined on a grid of ${\sim} 1\arcmin \times 1\arcmin$.
The NIR illumination correction is designed to remove the non-uniform illumination of the LED flats across the entire detector array. For VIS, it additionally corrects for a gradient due to time taken to open and close the shutter.

The NIR illumination correction reveals structure at the level of ${\sim} 5\%$ on sub-detector scales.
This is thought to be due to additive, low-level, diffuse charge persistence.
Nevertheless, the Q1 SGS NIR pipeline still applies this correction to the data multiplicatively.

For nominal-duration (\qty{566}{s}) VIS exposures, the illumination correction shows that large-scale illumination pattern and shutter effect are negligible. Instead, there are variations between and within detector quadrants, on the level of ${\sim} 2\%$. These corrections are applied multiplicatively, which is likely to be the best approach for improving the photometry of compact sources, but not necessarily for avoiding background artefacts.

We might therefore anticipate some residual signatures to be present in the background of the science images, particularly those of NIR.
In the SGS pipeline, these are easily dealt with by a small-scale background correction.
Since we wish to avoid this, we investigate the presence of residual background features.
If these are due to an interaction between the calibration and the background, we might expect them to be present in every image, and to vary little over sequences of images taken close in both time and sky position. If this is the case, then we may be able to remove these features from the data.

To robustly estimate the average background for a given band, we compute the median (after masking sources) across all exposures taken with that band in a group of $N_\mathrm{obs}$ neighbouring observations in the observation sequence.
The group comprises the $N_\mathrm{obs}/2$ observations on either side of the target observation.
Taking the median of sequential observations is a common approach for removing the time-varying background in ground-based infrared observations (e.g., \citealt{1995MNRAS.273..513B}). The technique has also been applied in visible bands, for example in CFHT deep imaging \citep{2012ApJS..200....4F,2015MNRAS.446..120D}.
In Q1, each of the three deep fields is covered by a distinct sequence of observation IDs.
To avoid grouping observations taken at very different times and sky positions, we truncate the group at the start and end of the observation ID sequence for each field.\footnote{For subsequent data releases, which will not contain distinctly separated fields, we will determine the time and angular separation between neighbouring observations and choose appropriate thresholds to achieve the same effect.}
Where possible, the number of observations on the opposite side of the target observation is increased, to maintain a total group size of $N_\mathrm{obs}$ observations.
The target observation itself is always excluded from the group to avoid any risk of removing true signal.
For NIR we use $N_\mathrm{obs}=6$, while for VIS we use $N_\mathrm{obs}=10$. These choices result in a correction that does not introduce significant noise when applied, while avoiding excessive computational demands.

In each ROS, the four exposures for each band are dithered by ${\sim}\qty{2}{arcmin}$, while each observation in the EWS is at least $\qty{20}{arcmin}$ from any other. Each background image is thus created from the median of many well-dithered images and is therefore free of contamination by sources.\footnote{Future application of this technique beyond Q1, to objects larger than the \Euclid FoV, may require the $N_\mathrm{obs}$ observations to be sampled from a wider area.}

We do indeed find that the resulting background images have a similar appearance across the whole Q1 data set, and vary little for neighbouring observations. To distinguish from other types of background (discussed in subsequent subsections), in this paper we always refer to this as the atemporal (i.e. constant, or at least slowly-varying) background.

Taking the median across many sequences of large images is a very computationally-intensive operation.
To improve this, we note that (a) variations in the atemporal background images are generally smooth within each detector image, and (b) where `edges' do occur in the NIR images, they correspond to the $204 \times 204$ pixel grid adopted for the LE2 large-scale illumination correction.
We therefore create the atemporal background images in two stages.
First, we create a coarsened version of the background in each science exposure, by taking the median on a grid.
Then, we take the median across the coarsened images within a group of neighbouring observations, as described above.
Before performing the coarsening, we mask sources: firstly, generous regions around bright stars, then all additional significantly detected source pixels (including bright persistence features). This ensures that (a) coarse bins with a significant fraction of their pixels covered by sources still accurately reflect the background level, and (b) those coarse bins that are entirely covered by sources are themselves masked for the second stage.

For NIR, we choose to coarsen each detector image in boxes of $51 \times 51$ pixels, to subsample the illumination correction grid by a factor of four.
Edge features due to the illumination correction can therefore be fully captured by the coarsened images.
For VIS, we choose boxes of $149 \times 149$ pixels, to obtain a similar angular scale as for NIR, without leaving narrow strips along any edge. This is appropriate, since the grid used for the VIS LE2 illumination correction roughly matches that used for NIR.
As well as being computationally convenient, this masked median-in-boxes coarsening helps remove small-scale persistence features, which are dealt with later (\refapp{persistence}), and further improves the robustness of the final background images.

We are interested in variations in the background across the detector array, but within a group of observations, there may be variations in the overall background level. To remove any such variations, we first subtract the median from each of the coarsened images, before taking the median of each pixel over the group of images. We then restore the background level in the final atemporal background image, by adding the median background level across the group of $N_\mathrm{obs}$ images (a single value). This is necessary to avoid a gradient being introduced when resampling the processed images, as explained further in \refapp{resampling}.

We construct an atemporal background image for each observation. These are `uncoarsened' back to the pixel scale of the original images, by linear interpolation for VIS and nearest-pixel interpolation for NIR to preserve the grid structure, then subtracted from the original images, as the first step of our post-processing.

Example atemporal background images are shown for NIR (\reffig{autodark_NIR}) and VIS (\reffig{autodark_VIS}). We see variations relative to the median sky level of roughly 10\% for VIS and 20\% for NIR. A comparison of the top two panels of \reffig{single_chip_processing} demonstrates the effect of subtracting the atemporal background from a NIR detector image.

\subsection{\label{ap:tartan}Tartan (NIR)}

\begin{figure*}[tbp]
\centering
\includegraphics[width=\hsize]{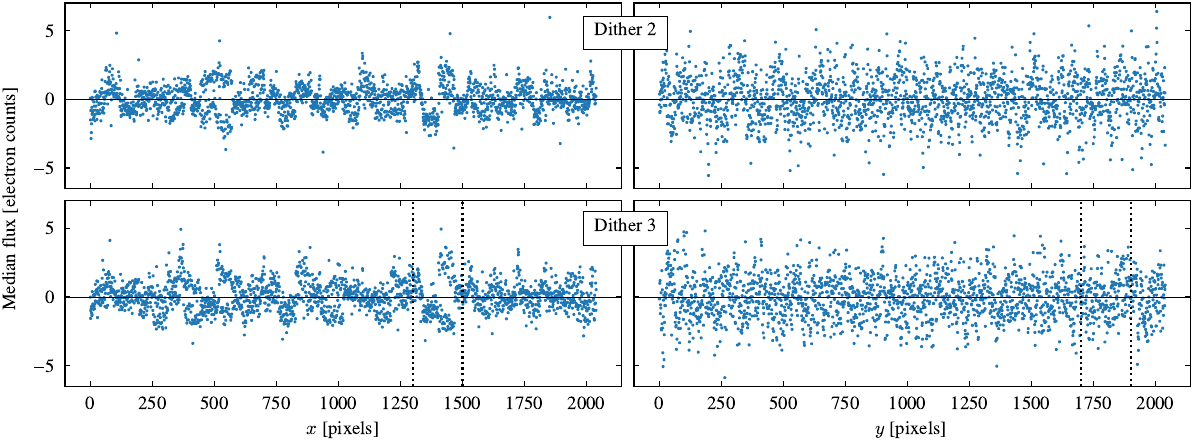}
\caption{\label{fig:tartan_projections}Background level of a NIR detector image, averaged over (\emph{left}) rows and (\emph{right}) columns for two exposures (dithers 2 and 3) within observation 2683. Bands of alternating column levels are apparent as bifurcations in the \emph{left} panels, in addition to other small-scale variations in both the columns and rows. Although some features appear to persist for a while, they typically vary between exposures. The \emph{bottom} row (dither 3) corresponds to the exposure shown in Figs.~\ref{fig:single_chip_processing} and \ref{fig:tartan_example}. The vertical dashed lines here indicate the region outlined in white in Figs.~\ref{fig:single_chip_processing}\,(c) and \ref{fig:tartan_example}.}
\end{figure*}

\begin{figure*}[tbp]
\centering
\includegraphics[width=\hsize]{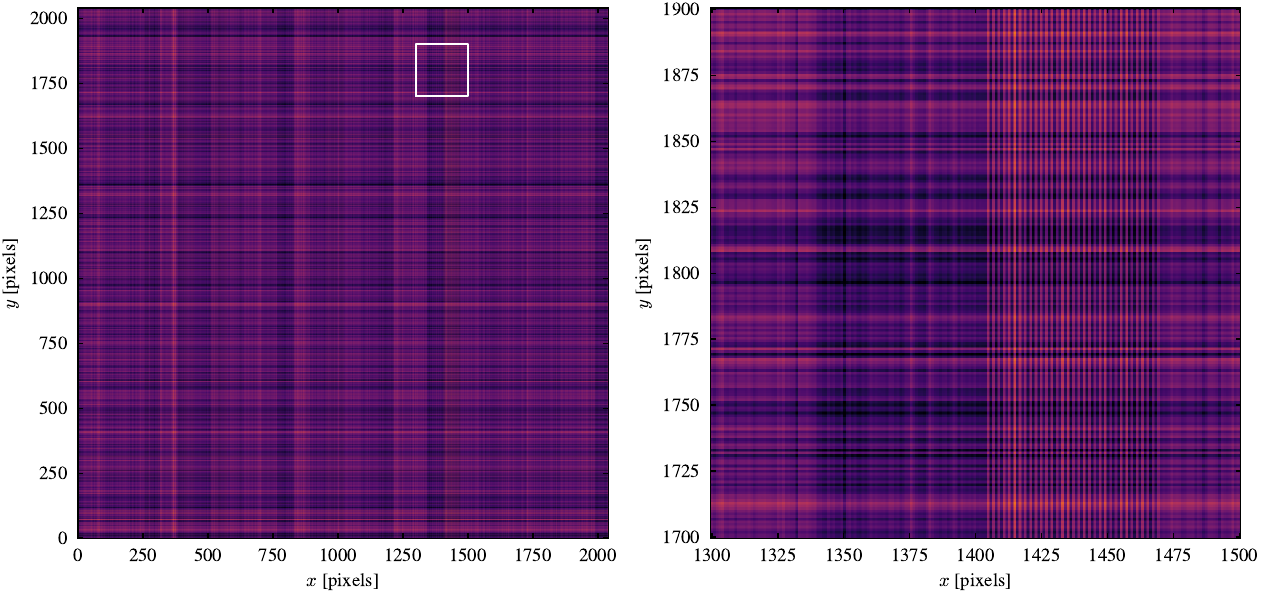}
\caption{\label{fig:tartan_example}Tartan correction for the same single detector shown in \reffig{single_chip_processing}, which is constructed from the 1D profiles plotted in \reffig{tartan_projections}. The \emph{left} panel shows the entire chip, while the \emph{right} panel shows a zoom-in of the region outlined in white. The same region is indicated in Figs.~\ref{fig:single_chip_processing}\,(c) and \ref{fig:tartan_projections}. We note the band of alternating column background levels from $x\approx 1405$ to $1470$. The colour scale is the same as in \reffig{single_chip_processing}.}
\end{figure*}

Close inspection of the NIR LE2 calibrated detector images reveals a subtle variation in the background level across entire rows and columns. This is particularly apparent once other post-processing steps have been applied; see, for example, \reffig{single_chip_processing}\,(c).
For simplicity -- given their combined appearance -- we refer to these features as `tartan'.\footnote{These features likely arise from readout correlations, also referred to as cross-talk or $1/f$ noise \citep{Jimenez21}.}
The pattern varies between exposures, and is therefore not captured by the atemporal background.
To construct a model of these features we take the average background level separately for each column and each row. In doing so, we must avoid contamination from sources. To remove relatively compact sources (significantly smaller than a detector) we define a mask containing pixels that are significantly detected above the background noise. We must also avoid including genuine extended, low-surface-brightness flux in our tartan model. The banding features have typical widths around 100~pixels or narrower. We therefore apply a circular median filter of diameter 200~pixels to the masked image and then subtract the resulting image from the original. This effectively removes any background variations larger than the tartan features. We then apply the mask to this image and take the median along each row and column, which avoids the background level from being biased by smaller-scale sources.

\refFig{tartan_projections} shows the resulting background level across the columns and rows of an example detector. The columns display a distinctive feature of bands in which the background level alternates from one column to the next. There are also smoother variations in background level on a scale of several rows or columns. The patterns in the two directions are coadded to produce a 2D image. We subtract the mean from this image to ensure the resulting correction does not change the average level of the background. An example of the tartan model is shown in \reffig{tartan_example}. This model is then subtracted from the original LE2 calibrated detector image. The resulting image has a notably smoother background, compare panels (a) and (c) of \reffig{single_chip_processing}.

\subsection{\label{ap:persistence}Persistence (NIR)}

\begin{figure*}[tbp]
\centering
\includegraphics[width=\hsize]{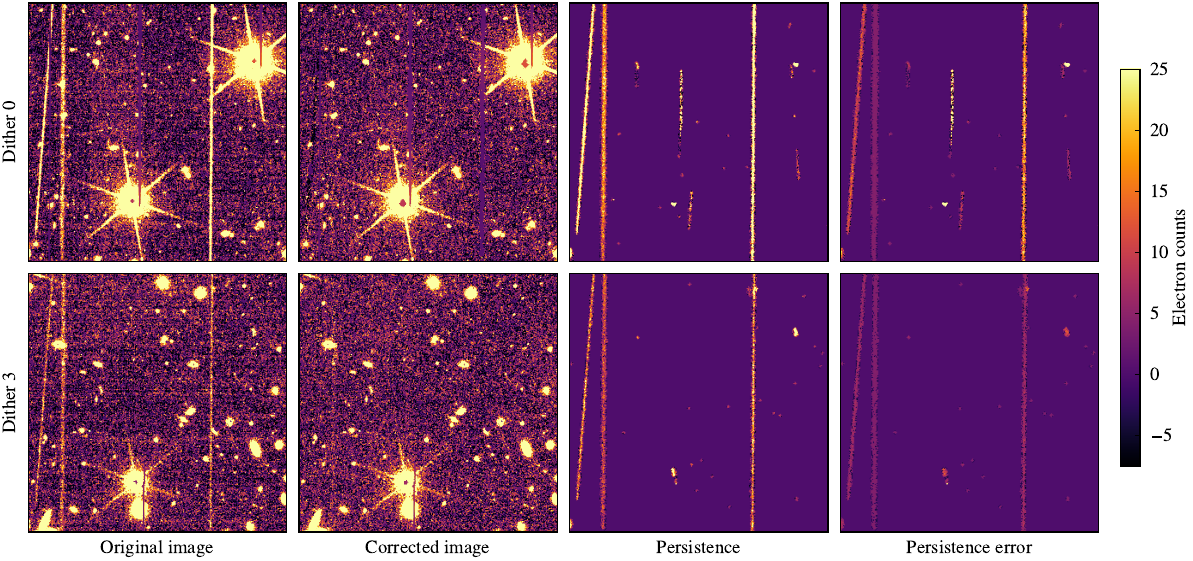}
\caption{\label{fig:persistence_zoom}Small region of a NIR detector image, indicated in \reffig{single_chip_processing}\,(d), illustrating the persistence correction. The rows display the first (\emph{top}) and last (\emph{bottom}) \HE band dither within the observation 2683; the last dither corresponds to the exposure shown in \reffig{single_chip_processing}. The columns show the image before (\emph{first column}) and after (\emph{second column}) the persistence correction, the persistence estimate itself (\emph{third column}), and its uncertainty (\emph{fourth column}).
The three almost vertical streaks are persistence features introduced by the spectroscopic infrared (SIR) exposures in a previous observation. These are well-modelled by the persistence image and leave behind very few residual artefacts.
Masked pixels have been replaced with the background level. The invalid-pixel mask, supplied with the LE2 calibrated frames, has been applied to the images in the left two columns. This has masked the persistence features associated with the two bright stars in the first dither, which were created in the preceding SIR exposure. In the corrected image, we additionally mask pixels where the uncertainty on the correction is $>20\,\mathrm{e}^-$. One of the streaks in the first dither reaches this threshold. Finally, we note that compact persistence features are introduced by undispersed exposures. There is one associated with each of the bright stars in the first dither, which was introduced by the preceding \JE band exposure. These features can still be seen (and corrected) in the last dither, where they are now slightly elongated due to small offsets between the consecutive \JE, \HE, and \YE exposures.}
\end{figure*}

Charge persistence is an effect whereby the observation of a bright source produces an artefact in subsequent exposures \citep{2008SPIE.7021E..0JS,2014PASP..126..739R,2024SPIE13103E..15K}. The most obvious persistence features in \Euclid NIR images are near-vertical streaks, resulting from bright sources in the grism-dispersed (SIR) images obtained by NISP, which are interleaved with the dithered NIR exposures in \JE, \HE, then \YE. The appearance of the same bright sources in the undispersed exposures also produces more compact, roughly circular, persistence features. The persistence in VIS images is negligible.

There has been extensive work on measuring and modelling the persistence behaviour of the NISP detectors \citep{2024SPIE13103E..15K} and \citetalias{EROData}. However, this previous work has taken an approach in which the persistence signal is predicted from the sources measured in previous exposures, assuming a model with parameters estimated from dedicated sequences of calibration exposures. Pixels that are substantially affected by persistence may be masked, while lower levels of contamination may be subtracted.
In the Q1 SGS NIR pipeline, only the previous spectroscopic exposure is considered, and pixels are masked if the predicted persistence signal is greater than $2\,\mathrm{e}^-$. This only removes a small fraction of the persistence features. The ERO pipeline used all exposures taken in the previous hour, and subtracted the predicted signal where it was less than $70\,\mathrm{e}^-$ (stronger features were left to be removed by outlier rejection during combining; see \citetalias{EROData}). This was more successful, but some issues remain. One problem is that the parameters of the persistence model have been found to vary significantly over time. The necessary frequent calibration sequences are not available for existing observations, and would be expensive to include in the future mission plan. \citetalias{EROData} recalibrated the model using observations, but the accuracy of this is limited. Another issue is that many of the persistence features in an image result from very bright sources observed several hours earlier. The model is not calibrated for source flux levels beyond saturation, and in any case the true flux of these sources is not well known, so their subsequent persistence signal cannot be accurately modelled.

We take an alternative approach, attempting to determine the persistence signal in a given exposure from the persistence features in temporally-neighbouring exposures. Consistent with the rest of our post-processing, we only use the LE2 calibrated NIR and SIR science products, and thus do not directly use any calibrations. Our method uses the idea that a persistence signal will be present in each exposure following the observation of a bright source, but with some decay over time. The sky sources move between dithers, so taking a low quantile (between the minimum and median) over a stack of sequential dithers (which may span multiple observations) around a target exposure, produces an estimate of the persistence signal in the target image. When building the sequence of images for a given target exposure, we only include exposures in the same band. Each \JE, \HE, \YE sequence is taken with the same pointing, so they do not help with rejecting sky sources. We use a sequence of nine exposures: four exposures either side of the target, plus the target itself. The sequence therefore extends into the ROS before and after the target. We apply the atemporal background and an initial tartan correction to these exposures, to avoid including detector-scale background variations and improve the quality of the persistence estimate.

If no persistence features were created during the exposure sequence, then the above approach would suffice. However, if a feature is created during the sequence, then the quantile flux may occur before the persistence appeared, and hence does not provide any useful information. Conversely, if a persistence feature does not appear until after the target image, then we do not want to include it in our correction. We therefore need to mask pixels in the stack depending on when the persistence feature is created.
To do so, we identify sources likely to introduce a multi-exposure persistence feature, by thresholding all images taken during the three ROSs included in the sequence (NIR and SIR), at $\num{10000}\,\mathrm{e}^-\,\unit{pixel^{-1}}$. This threshold corresponds to a persistence signal of roughly $\num{20}\,\mathrm{e}^-\,\unit{pixel^{-1}}$ in the immediately following exposure \citepalias{EROData}. While sources up to an order of magnitude fainter produce significant persistence signals, these rapidly decay. Inspection reveals that only sources with a signal greater than $\num{10000}\,\mathrm{e}^-\,\unit{pixel^{-1}}$ result in a detectable persistence signal in the subsequent exposure. We record the latest time such a persistence-causing signal is flagged for each pixel. If that time is equal or prior to the time of the target image, then we mask pixels earlier in the stack. If it is after the target image, then we mask pixels later in the stack.

The quantile we use is the second lowest value over the masked sequence of exposures. We initially used the minimum but found it less robust. Taking the median might seem like a more obvious choice, but it is more likely to be contaminated by a sky source for a short sequence of exposures.
We completely mask any pixels where there are fewer than four unmasked values remaining in the sequence, since the chosen quantile is likely to be contaminated by sky sources.

At this point we have, for each pixel, an estimate for the persistence flux, the time since the last persistence feature was created (if it occurred within the sequence), and the time between the target exposure (which we want to correct) and our persistence estimate.
The persistence estimate is lower than the true persistence in the target image, since the persistence signal has decayed in the intervening time. Simply subtracting the persistence estimate from the target image actually works well for old persistence features (e.g., formed by very bright sources before the previous ROS). For more recent features, this approach would still be an improvement, but would not remove all of the persistence signal. However, if we can model the decay, then we can rescale the persistence estimate to the time of the target image.

To determine a suitable model, we combine all the persistence estimates and run source detection to find the bright features. We then plot the logarithmic flux of each feature against the time since the persistence feature was created. These results show that the flux decay is well modelled by a power law, as found by \citet{2024SPIE13103E..15K} and \citetalias{EROData}. For a persistence signal of $f_0$ estimated at time $t_0$ after the feature was introduced, the signal in the target exposure, $f_1$, at time $t_1$, is modelled by
\begin{equation}
f_1 = f_0\, ({t_0}/{t_1})^a \,.
\end{equation}
The decay index, $a$, varies somewhat by feature and detector, likely reflecting a dependence on detector, position on the detector, brightness of the originating source, etc. as expected \citep{2024SPIE13103E..15K}. In principle, one could attempt to include this information in the model. However, here we simply take the median decay index of $a = 1.0$. Individual values vary about this fiducial slope with a scatter of $\sigma_a \sim 0.2$. Since the persistence estimate is close in time to the target, the decay correction factor is small, and hence this simplistic model works rather well.
After applying the scaling to reverse the decay, we have an improved persistence estimate for each target image. This persistence estimate is subtracted from the original target image, followed by a fresh estimate of the atemporal background and tartan pattern, to reduce contamination of these final corrections by persistence features.

Next, we create an estimate of the uncertainty introduced by the persistence correction. This includes the uncertainty on the original persistence estimate (from the RMS image), the decay slope $\sigma_a$, and the time the persistence feature formed. Since the persistence feature was formed over the course of an exposure, we assume that $t_1$ has an uncertainty corresponding to a quarter of the exposure time, that is, \qty{22}{seconds}.
The estimated uncertainty on the correction is added in quadrature to the original RMS image.

Finally, very strong persistence features are difficult to correct, due to both the uncertainty of the decay scaling and the initial persistence estimate. Even a perfect correction would leave strong residuals, due to the Poisson error on the persistence signal in the target exposure itself. We therefore mask pixels for which the uncertainty on the correction is greater than $20\,\mathrm{e}^-$.

An illustrative example of the correction is shown in \reffig{persistence_zoom}. The effect of the correction for a single dither may be judged by comparing panels (a) and (d) of \reffig{single_chip_processing}. Finally, the benefits of the correction after combining multiple dithers is apparent in \reffig{nicl_vs_MER}. The persistence features introduced by previous SIR exposures are visible as prominent diagonal streaks in the LE2 and MER products, but are absent in our post-processed image.

\subsection{\label{ap:resampling}Image resampling (NIR \& VIS)}

The focus of our post-processing is to produce images centred on a set of specific targets (galaxy clusters). We specify the position and size of each target region, then combine, for each band, all dithers that overlap with it.
There are usually at least four dithers covering a given target region. For our ICL measurements, we typically produce a $30\arcmin\times30\arcmin$ image around each cluster, to allow for a background subtraction on sufficiently large scales (described in \refapp{background_subtraction}).

As explained later, in \refapp{matching}, it is helpful to include data from neighbouring exposures to constrain the background matching correction. We therefore search for dithers that overlap the region formed by expanding the target area by one dither offset on each side. Within these dithers, we identify the individual detector images that overlap with the expanded region. We then use \swarp{} \citep{Bertin+2002} to resample each detector image onto a common pixel grid centred on the target, using the \texttt{LANCZOS2} interpolation kernel. We do not yet combine the resampled images, since we first apply a background matching procedure. We disable \swarp{} cropping of detector images which partially overlap the target region, to retain more pixels for this correction.

Geometric distortions due to the camera optics produce a small, smooth variation in the solid angle sampled by each pixel in the image (at a level of ${\sim} 1\%$ for NIR and ${\sim} 2\%$ for VIS). The atemporal background correction removes any resulting gradients, along with the other background features visible in Figs.~\ref{fig:autodark_NIR} and \ref{fig:autodark_VIS}, to produce a flat background in each detector image. However, the geometric distortions are accounted for by \swarp{}, so an initially-flat, non-zero flux level would result in a gradient after resampling. This motivates restoring the average background level in the atemporal background estimate (as described in \refapp{autodark}). After subtracting the atemporal background estimate, the images have a nearly-zero background level prior to resampling, which eliminates the introduction of a gradient.

We also produce resampled weight images using the inverse variance maps supplied with the LE2 calibrated data. The weights are set to zero for bad pixels. These include pixels marked as invalid by the quality flags accompanying the LE2 data as well as those masked by our persistence correction. When resampling with \swarp{}, if one pixel within the interpolation kernel has a weight of zero, the resulting pixel in the resampled image will also have a weight of zero. A consequence of choosing the \texttt{LANCZOS2} kernel is, therefore, that the bad pixel footprint is broadened in the NIR images: a single bad pixel becomes around 16 pixels. Larger interpolation kernels would be even more problematic. As a result, the final combined images still contain a significant number of invalid, and hence masked, pixels, despite the multiple overlapping exposures. This is unfortunate from a cosmetic perspective, but is a conservative choice for our science. For VIS this is not a significant issue, since we downsample the pixel resolution by a factor of three.

For our ICL measurements we resample to a pixel scale of \qty{0.3}{arcsec.pixel^{-1}}. This choice makes the undersampling of VIS images worse, but since our intended science targets are well resolved, we can afford to trade some spatial resolution for processing speed. A unified pixel scale is also convenient to ensure identical apertures when measuring the colour profiles of the ICL (\refsec{measurement}).\footnote{The images could easily be produced at the full VIS resolution if required.} All images are scaled to share the same magnitude zero point of \qty{23.9}{mag}, such that one ADU corresponds to one \unit{\micro Jy}.
This step results in resampled detector images and their weights, with pixels aligned and ready for differencing (\refapp{matching}) then combining (\refapp{combining}).

\subsection{\label{ap:matching}Background matching (NIR \& VIS)}

The atemporal background correction successfully removes most of the background systematics. However, we find that there are still low-level discontinuities in the background of combined images.

We therefore apply a procedure to correct the residual mismatch in the background level of adjacent detector images. Such a `continuity correction' is not unique to \Euclid cameras. Where individual detectors contain sufficient uncontaminated background pixels, modelling and subtraction of the background level -- including careful, often iterative, masking -- can be very successful \citep{2021A&A...654A..40T,2022ApJ...940L..51M}. However, where sources may be larger than an individual detector, an alternative approach is required. For example, \citet{Waters+2020} achieved good results for individual exposures of the Pan-STARRS gigapixel camera by applying a constant additive correction to each readout unit to minimise the difference in the background levels across adjacent edges. However, gaps between the detectors and bright/extended sources can hinder accurate inference of the corrections. If the dither pattern permits, one may make use of the overlaps between the dithers to circumvent these issues \citep{2021MNRAS.507..963G}. Thanks to the observational strategy of the \Euclid ROS, we can take this approach, which allows us to remove detector discontinuities even in the presence of very extended sources.

Essentially, the goal of matching two adjacent detector images in one exposure is addressed by matching each to another detector image (from another exposure) that overlaps with both. Since extended or crowded sources may be present, it is more robust to compute the background difference between two overlapping detector images by subtracting one from the other, rather than attempting to measure the background levels in the respective overlap regions. Computing such difference images is convenient using the resampled images created prior to combining.

To start, a complete list of unique pairs of overlapping detector images is compiled. We then compute the difference image for each overlap region. A weight image is also formed from the inverse of the sum of the corresponding variance maps. Although source fluxes are mostly cancelled in the difference image, we find significant systematic residuals from bright stars. To mitigate their impact, we create source masks from the detector images and set weights to zero for pixels in the difference image that overlap with the source masks. The background offset between the $i$th and $j$th detector images is estimated by taking the weighted mean $\Delta_{ij}$ of the difference between the two images. We seek to determine a constant correction $C_k$ for each detector, which minimises the $\Delta_{ij}$. We therefore construct a linear system of equations,
\begin{equation}\label{eq:bkg_match}
C_j - C_i = \Delta_{ij}\quad \text{for all overlapping }i,j\,.
\end{equation}
The number of unknowns equals the number of detector images that overlap with at least another image. There are thus more equations than unknowns because each detector image often overlaps with several others. We divide Eq.~\eqref{eq:bkg_match} by the propagated uncertainty on $\Delta_{ij}$, to down-weight overlaps with a small number of pixels. We then determine the corrections, $C_k$, using a linear algebra least-squares solver.

The system has a rank deficiency: the solution is not unique, because it cannot constrain the overall background level across each complete set of overlapping images. In our case, the \Euclid survey strategy ensures there is only ever one degree of freedom. We adopt the solution with the smallest Euclidean norm, such that the mean correction is zero. Since the original images have each had their expected background level subtracted by the atemporal background correction, after applying the correction $C_k$, the images have a global background level of zero.

Each of the processing steps above helps to improve the quality of the background matching. Background artefacts that differ between the images in an overlapping pair would obviously introduce noise into the correction, resulting in spurious offsets in the level of adjacent detector images. Because of this, correcting relatively small-scale artefacts, such as persistence features with a width of \qty{1}{arcsec}, helps to improve the flatness of the background on scales exceeding \qty{10}{arcmin}.

\subsection{\label{ap:combining}Combining images (NIR \& VIS)}

We use \swarp{} to combine the resampled, background-matched detector images. We choose the \texttt{CLIPPED} combine mode, with default parameters, which computes the weighted mean after clipping pixels that vary significantly from the median of the exposures. This mode helps reduce residual artefacts, such as cosmic rays and detector issues. Moreover, the inverse-variance weighted mean is optimal for achieving the best \snr. The \texttt{MEDIAN} combine mode, as adopted for Q1 MER stacks \citep{Q1-TP004}, may better mitigate impacts of the artefacts, but would be noisier than the weighted mean combination and therefore not ideal for LSB science. This time we allow \swarp{} to crop detector images so the dimensions of the combined image are exactly the size requested.

At this stage, we have produced science-ready combined images for each galaxy cluster. Applying the full post-processing to all observations in Q1 takes only a few days, utilising 16 cores of a modern server. Since the task can be parallelised at the observation level, it is easy to accelerate with high performance computing. Importantly, so far, we have not subtracted any local estimate of the background in the target image. We perform a final background subtraction, optimised for the angular size of each galaxy cluster, as part of our ICL measurement procedure, described in \refsec{measurement}.

\section{\label{ap:measurement}Measurement procedure details}

\subsection{\label{ap:background_subtraction}Background estimation and subtraction}

We use a consistent approach to background estimation throughout this work, varying the spatial scale to suit the object under study. We follow the approach popularised by \texttt{SExtractor} \citep{1996A&AS..117..393B}, but use the \texttt{photutils.Background2D} implementation \citep{photutils}. The spatial scale is quantified by the box size of a square grid. In each box, the background level and RMS noise are estimated by the sigma-clipped median and standard deviation, respectively. This is followed by a $3 \times 3$ median filter over the boxes, then bicubic interpolation back to the original pixel grid. The resulting background level is then typically subtracted from the original image, while the noise is used to set the threshold for detecting sources.

When constructing masks (see \refapp{masking}) we choose relatively small background box sizes, to ensure we detect genuine sources. However, when performing photometry on our cluster targets we take a conservative approach, to ensure an accurate background estimate and avoid subtracting ICL flux. Although the background estimates use robust statistics, we also apply a mask to exclude flux from all detected sources, and hence reduce any potential biases. We estimate the background in the cluster image using a box size corresponding to \qty{1}{Mpc} at the redshift of the target cluster. This is roughly the virial radius of a $10^{14}\Msun$ cluster, and well beyond where we expect to measure ICL for individual objects in the EWS. At $z=0.1$, for example, this equates to a box size of \qty{9}{arcmin} or \qty{1800}{pixels}. We have tested that this choice of background scale does not affect our measured ICL profiles, using both real and mock clusters (see \reffig{mock_test}). For each cluster, we produce images with a fixed $30\arcmin \times 30\arcmin$ size, centred on the BCG. This provides sufficient coverage to contain a $3 \times 3$ box grid for our largest targets in Q1.

\subsection{\label{ap:masking}Masking}

We need to construct masks for use when measuring the ICL photometry and estimating the background. The first (the `measurement mask') needs to mask all sources, with the exception of the BCG and ICL. The second (the `background mask') needs to mask the same sources, but additionally must conservatively mask the region where the BCG and ICL are significant.

A mask of the region containing significant BCG+ICL signal is needed to construct both the measurement and background masks, so we describe the creation of this first. This `ICL mask' is constructed through an adaptive smoothing algorithm that seeks to identify the extended diffuse light around the cluster centre. This begins by estimating the background noise with a 300 pixel box size. The image is smoothed on a scale $\sigma_\mathrm{s}$ by a Gaussian median filter\footnote{We have developed a fast, circular, centrally-weighted median filter. This takes the median over pixels sparsely sampled using weights from a Gaussian kernel with a cut-off at $5\sigma_\mathrm{s}$. For large kernels, the data is first coarsened by taking the median in square blocks of pixels. This combines benefits of Gaussian and median filtering, with a computation time that only grows linearly with the filter size.} and sources are detected using a threshold of twice the background noise. The smoothing scale is initially set to $\sigma_\mathrm{s} = \qty{20}{pixels}$, then iteratively adjusted through a binary search: if no source is detected at the position of the cluster centre, then the smoothing scale is much larger than the ICL and so $\sigma_\mathrm{s}$ is reduced; if a source is detected, the smoothing is increased. This continues until an optimal $\sigma_\mathrm{s}$ is found (within a relative accuracy of 5\%), which maximises the smoothing, while still permitting detection of the central source. The central source is then dilated using a circular structure with radius equal to the final $\sigma_\mathrm{s}$. The resulting ICL mask approximately encompasses the extent of the diffuse light in the cluster.

The measurement mask is formed from the union of three masks: a `bad pixel mask', a `regular objects mask', and a `central objects mask'. The bad pixel mask simply identifies any pixels without valid data after combining the individual exposures (\refapp{combining}). To create the regular objects mask, we first estimate the background level and noise with a box size of \qty{500}{pixels}. The image is smoothed with a small Gaussian kernel ($\sigma = \qty{1}{pixel}$), then sources are detected above a threshold of three times the background noise and requiring at least 5 detected pixels per source. The detected sources are then deblended. Sources embedded in the ICL will appear larger than they otherwise would, resulting in us masking much of the signal we are trying to detect. We therefore exclude any sources that overlap with the ICL mask, which will instead be dealt with by the central object mask. The remaining detected sources are then grown by iteratively dilating until they are 50\% larger than their original size. This mask almost completely covers sources that lie beyond the central ICL region (the performance of our masking is demonstrated in \refsec{simulations}).

The central objects mask is created specifically for sources located within the ICL region that were excluded from the regular object detection. We first determine a local background level by filtering the image using a Gaussian median filter with a smoothing scale of $0.25\sigma_\mathrm{s}$, where $\sigma_\mathrm{s}$ is the optimal value determined when constructing the ICL mask. We use the same Gaussian median filter applied to the absolute residuals to estimate the RMS noise. Sources are detected using a threshold of twice the noise above the local background, then more aggressively deblended. The central source (the BCG) is excluded, and the detections are restricted to a region formed by dilating the ICL mask to twice its original size. The remaining sources are then grown by iteratively dilating until they are 25\% larger than their original size. Our tests determined that this approach provides adequate masking while avoiding over-masking of genuine ICL flux (discussed further in \refsec{simulations}).

Finally, the background mask is formed from the union of the measurement mask and the ICL mask. It should be noted that, while masking is an important element of measuring the ICL and the background, the detailed choices are not critical. Our measurements use robust statistics, which are insensitive to contamination by source flux that escapes our mask, so long as the fraction of affected pixels is relatively small.

To create the VIS mask, we apply the above steps directly to the processed VIS image. However, for the NIR mask, we first create a combined \YJH image, then apply the above steps to produce a single \YJH mask that is used for each NIR band. This helps ensure that faint sources are included and consistently masked across the NIR bands. To create the combined \YJH image, we first subtract the background from the image for each band individually. For this, we generate and apply a background mask as described above, then subtract a background estimated using a box size of \qty{300}{pixels} and a median filter size of \qty{3}{pixels}. The three background-subtracted images are then weighted by their bandwidth in terms of frequency (weights of $0.375$, $0.358$, and $0.267$ for \YE, \JE, and \HE), and summed to mimic an image with a transmission window extending from \qty{950}{nm} to \qty{2220}{nm}.

\subsection{\label{ap:photometry}Photometry with elliptical isophotes}

We perform photometry on our fully-processed, background-subtracted cluster images, masking all sources other than the BCG and ICL, as described above. The images are centred on the specified central coordinate of the cluster. How this centre is defined is described in the individual science papers (see \citetalias{Q1-SP007}; \citetalias{Q1-SP008}; G26), but typically the BCG is identified and assumed to lie at the centre of the ICL. Elliptical isophotes are then fit to the image using \texttt{AutoProf} \citep{2021MNRAS.508.1870S}.

\texttt{AutoProf} is designed to handle many aspects of the measurement process, but we primarily use it to fit the shapes of the isophotes. We hold the centres and semi-major axes of the isophotes fixed, but allow \texttt{AutoProf} to iteratively vary the ellipticity and position angle (PA) of each isophote to best match the intensity distribution. The semi-major axes are defined starting at 1 pixel, with each successive ellipse being 40\% larger than the last. As a result, each annulus has a semi-major axis width of $r_\textrm{c}/3$, where $r_\textrm{c}$ is the semi-major axis at the centre of the annulus.

Since we have performed a careful background subtraction, we do not allow \texttt{AutoProf} to re-estimate the background level. \texttt{AutoProf} incorporates a number of elements to ensure the robustness of the isophotes it determines. These include clipping the bright pixels, including those from sources that have escaped masking, and regularisation of the isophote parameters, by penalising substantial changes in ellipticity and PA between adjacent isophotes. \texttt{AutoProf} also uses a signal-to-noise criterion to determine the maximum semi-major axis at which it can measure the isophotal shape. Beyond this, the shape of the ellipses is no longer allowed to vary. As we are interested in the shape of the ICL at large radii \citepalias{Q1-SP007}, we lower this signal-to-noise threshold from the default value of 2 to 1. Except where mentioned previously, we leave all other parameters controlling the isophote fitting at their default values.

Although \texttt{AutoProf} can measure the surface brightness in each annulus, we implement our own measurements for flexibility, using \texttt{photutils.aperture}. For each annulus (the region between two adjacent isophotes), the measurement mask is applied to exclude contaminating sources, and the median flux per pixel is calculated using iterative $3\sigma$-clipping to reject any residual contamination from imperfectly masked sources or artefacts. This provides a very robust measure of the intensity corresponding to each isophote. We also measure the standard error of the masked, sigma-clipped pixels, to use as an estimate of the statistical uncertainty on the measured intensity at each isophote.

Some previous studies use circular annuli to measure the surface-brightness profiles of the BCG+ICL (e.g., \citealt{2018MNRAS.474.3009D,2019ApJ...874..165Z,2023MNRAS.521..478G}). However, these objects typically display significant ellipticity \citepalias{Q1-SP007}, and hence the flux around a circular aperture will vary substantially. As a result, sigma-clipping is no longer effective for removing contaminating sources. More importantly, this means that masking pixels in different azimuthal regions of the annulus biases the average flux in the annulus. The measured profile thus becomes dependent on the position of masked sources. In our tests, we find this to be a significant source of noise, and it can also produce a systematic bias, for example, if the distribution of cluster galaxies correlates with the orientation of the ICL \citepalias{Q1-SP008}. The use of elliptical isophotes mostly avoids these issues. For comparison purposes, it is possible to convert the more reliable ellipse-based model to equivalent circular profiles (e.g., \citealt{2007ApJ...666..147G}).

When considering colour profiles (e.g., \citetalias{Q1-SP008}), it can be preferable to perform photometry in different bands using identical apertures. In addition to measuring each band independently, we therefore also fit elliptical isophotes to a combined \YJH image, then perform photometry in each separate band (including \IE) using the \YJH-defined ellipses and combined \YJH{}+\IE masks. The creation of the \YJH image is described at the end of \refapp{masking}.

\section{\label{ap:simulations}Construction of mock cluster images}

Our mock images are built directly from the Hydrangea cosmological hydrodynamical simulations \citep{2017MNRAS.470.4186B}, a set of zoom-in galaxy cluster simulations run within the EAGLE galaxy formation framework \citep{2015MNRAS.446..521S}. Following the approach of \citet{2022MNRAS.513.1459M}, each stellar particle is assigned a spectral energy distribution (SED) according to its current stellar mass, age, and metallicity. This SED is redshifted to the appropriate frame, using the epoch of the simulation snapshot, and multiplied by the \Euclid transmission function to obtain an apparent flux. To reduce pixel-scale artefacts in diffuse regions, the particle distribution is adaptively smoothed prior to projection. The projected images are then convolved with the \Euclid PSF (which includes the large-scale wings; \citetalias{EROData}) and rendered at a pixel scale of \qty{0.3}{arcsec.pixel^{-1}}.

At each of six target redshifts, $z \approx 0.1$, $0.2$, $0.4$, $0.54$, $0.7$, and $0.8$, we select 30 to 50 unique clusters spanning the mass range $M_{200} = 10^{14}\,M_{\odot}\text{ to }10^{15}\,M_{\odot}$. Many of these correspond to the same structures at different stages of their evolution. Each system is projected along the $x$, $y$, and $z$ axes, producing 776 mock cluster configurations. We use these for all the tests of our measurement methodology detailed below.

For each projected cluster, we construct two mock images without any background. The first image contains the stellar light only from the BCG+ICL component of the cluster. This `ideal' image serve as the truth against which we compare our measurements. The second additionally contains the stellar light from all satellite galaxies within $R_{200}$. The separation of these two components is achieved directly from the simulations. In Hydrangea, satellite substructures are first identified using \texttt{Subfind} \citep{2010MNRAS.405.1544D}, which detects friends-of-friends subhaloes and then locates gravitationally-bound overdensities within them. These detections are reprocessed (using \textsc{CANTOR}; Bah{\'e} et al. in prep.), to better identify those stellar particles bound to each subhalo. The resulting objects are classified as satellites; all remaining particles constitute the BCG+ICL.
\end{appendix}

\label{LastPage}
\end{document}